\documentclass[fleqn,usenatbib]{mnras}

\usepackage{newtxtext}
\usepackage[T1]{fontenc}

\DeclareRobustCommand{\VAN}[3]{#2}
\let\VANthebibliography\thebibliography
\def\thebibliography{\DeclareRobustCommand{\VAN}[3]{##3}\VANthebibliography}

\usepackage{graphicx}	% Including figure files
\usepackage{amsmath}	% Advanced maths commands
\usepackage{amssymb}	% Extra maths symbols
\usepackage{url}

\allowdisplaybreaks[1]
\usepackage{hyperref}
\newcommand{\VEC}{\mathbfit}
\newcommand{\MAT}{\mathbfit}
\newcommand{\hMpc}{\,h^{-1}\,{\rm Mpc}}
\newcommand{\oV}{\,(h^{-1}\,{\rm Mpc})^{-3}}

\newcommand{\hV}{\,(h^{-1}\,{\rm Gpc)^3}}

\title[First test of the consistency relation for LSS]{First test of the consistency relation for the large-scale structure using the anisotropic three-point correlation function of BOSS DR12 galaxies}

\author[N. S. Sugiyama et al.]{
Naonori S. Sugiyama$^{1}$\thanks{E-mail: nao.s.sugiyama@gmail.com}\\
$^{1}$ National Astronomical Observatory of Japan, Mitaka, Tokyo 181-8588, Japan\\
\newauthor 
Daisuke Yamauchi$^{2}$, Tsutomu Kobayashi$^{3}$, Tomohiro Fujita$^{4,5}$, Shun Arai$^{6}$, and Shin'ichi Hirano$^{7}$\\
$^{2}$ Department of Physics, Faculty of Science,
Okayama University of Science, 1-1 Ridaicho, Okayama, 700-0005, Japan\\
$^{3}$ Department of Physics, Rikkyo University, Toshima, Tokyo 171-8501, Japan\\
$^{4}$ Waseda Institute for Advanced Study, Shinjuku, Tokyo 169-8050, Japan\\
$^{5}$ Research Center for the Early Universe, The University of Tokyo, Bunkyo, Tokyo 113-0033, Japan\\
$^{6}$ Kobayashi-Maskawa Institute, Nagoya University, Nagoya 464-8602, Japan\\
$^{7}$Oyama National College of Technology, Oyama 323-0806, Japan\\
\newauthor 
Shun Saito$^{8,9}$, Florian Beutler$^{10}$, and Hee-Jong Seo$^{11,12,13}$\\
$^{8}$ Institute for Multi-messenger Astrophysics and Cosmology, Department of Physics,\\
Missouri University of Science and Technology, 1315 N. Pine St., Rolla MO 65409, USA\\
$^{9}$ Kavli Institute for the Physics and Mathematics of the Universe (WPI), \\
Todai Institutes for Advanced Study, The University of Tokyo, Chiba 277-8582, Japan\\
$^{10}$ Institute for Astronomy, University of Edinburgh, Royal Observatory, Blackford Hill, Edinburgh EH9 3HJ, UK\\
$^{11}$ Department of Physics and Astronomy, Ohio University, Clippinger Labs, Athens, OH 45701 \\
$^{12}$ Physics Division, Lawrence Berkeley National Laboratory, 1 Cyclotron Road, Berkeley, CA 94720, USA\\
$^{13}$ Berkeley Center for Cosmological Physics, Department of Physics, University of California, Berkeley, CA 94720, USA
}
\date{}

\pubyear{}

\begin{document}
\label{firstpage}
\pagerange{\pageref{firstpage}--\pageref{lastpage}}
\maketitle

\begin{abstract}
    
    We present, for the first time, an observational test of the consistency relation for the large-scale structure (LSS) of the Universe through a joint analysis of the anisotropic two- and three-point correlation functions (2PCF and 3PCF) of galaxies. We parameterise the breakdown of the LSS consistency relation in the squeezed limit by $E_{\rm s}$, which represents the ratio of the coefficients of the shift terms in the second-order density and velocity fluctuations. $E_{\rm s}\neq1$ is a sufficient condition under which the LSS consistency relation is violated. A novel aspect of this work is that we constrain $E_{\rm s}$ by obtaining information about the nonlinear velocity field from the quadrupole component of the 3PCF without taking the squeezed limit. Using the galaxy catalogues in the Baryon Oscillation Spectroscopic Survey (BOSS) Data Release 12, we obtain $E_{\rm s} = -0.92_{-3.26}^{+3.13}$, indicating that there is no violation of the LSS consistency relation in our analysis within the statistical errors. Our parameterisation is general enough that our constraint can be applied to a wide range of theories, such as multicomponent fluids, modified gravity theories, and their associated galaxy bias effects. Our analysis opens a new observational window to test the fundamental physics using the anisotropic higher-order correlation functions of galaxy clustering. 

\end{abstract}


%%%%%%%%%%%%%%%%%%%%%%%%%%%%%%%%%%%%%%%%%%%%%%%%%%

%%%%%%%%%%%%%%%%% BODY OF PAPER %%%%%%%%%%%%%%%%%%

\section{Introduction}
\label{Sec:Introduction}

The consistency relation of multi-point statistics in cosmology is a relation that non-perturbatively relates an $n$-point statistic of cosmic fluctuations to an $(n-1)$-point statistic. This relation holds in the limit that one of the $n\geq3$ wavenumbers is much smaller than the others, the so-called \emph{squeezed limit}. Originally proposed for single-field inflationary models~\citep{Maldacena:2002vr,Creminelli:2004yq}, a similar consistency relation was later invented in the large-scale structure (LSS) of the Universe~\citep{Peloso:2013zw,Kehagias:2013yd,Creminelli:2013mca}.

The LSS consistency relation is due to the fact that the equations for cosmic fluctuations are invariant under the Galilean transformation~\citep{Scoccimarro:1995if,Creminelli:2013mca}. In particular, the Galilean transformation eliminates the large-scale flow of matter at equal times, so that all higher-order nonlinear contributions beyond the leading order in perturbation theory are cancelled out when computing $n$-point statistics. This behaviour is called the equal-time consistency relation or infrared (IR) cancellation~\citep{Jain:1995kx,Scoccimarro:1995if,Kehagias:2013yd,Peloso:2013zw,Sugiyama:2013pwa,Sugiyama:2013gza,Blas:2013bpa,Blas:2015qsi,Lewandowski:2017kes}. On the other hand, various conditions have been proposed to violate this consistency relation, such as multicomponent fluids~\citep{Tseliakhovich:2010bj,Yoo:2011tq,Bernardeau:2011vy,Bernardeau:2012aq,Peloso:2013spa,Creminelli:2013poa,Lewandowski:2014rca,Slepian:2016weg}, primordial non-Gaussianity~\citep{Berezhiani:2014kga,Valageas:2016hhr,Esposito:2019jkb,Goldstein:2022PhRvD.106l3525G}, and violation of the equivalence principle~\citep{Creminelli:2014JCAP...06..009C,Inomata:2023arXiv230410559I}. It is therefore fundamental to test whether our Universe has a simple structure that satisfies the LSS consistency relation.

\citet{Crisostomi:2019vhj,Lewandowski:2019txi} pointed out that the Degenerate Higher-Order Scalar-Tensor (DHOST) theories~\citep[for reviews, see][]{Langlois:2018dxi,Kobayashi:2019hrl}, a type of modified gravity, also violate the LSS consistency relation. The reason is that when the second-order dark matter density fluctuations are decomposed into two independent components, the shift term and the tidal force term~\footnote{The scale dependence of the second-order density fluctuation of dark matter is generally decomposed into the growth, shift, and tidal terms~\citep{Schmittfull:2015PhRvD..91d3530S}. However, due to the condition that the ensemble average (infinite space integral) of the dark matter density fluctuation is zero, the coefficients of the three terms are related, and there are only two independent components. This relation is known to break down when galaxy bias effects are taken into account, in which case these three independent components must be considered~\citep[e.g.,][]{Desjacques:2016bnm}.}, DHOST theories change both terms from the values of general relativity (GR)~\citep{Hirano:2018uar}. On the other hand, Horndeski theories~\citep{Horndeski:1974wa,Deffayet:2011gz,Kobayashi:2011nu}, a subclass of DHOST theories, change only the tidal term from GR, leaving the shift term unchanged~\citep{Bernardeau:2011JCAP...06..019B,Takushima:2013foa,Bartolo:2013ws,Bellini:2015wfa,Burrage:2019afs}. Focusing on the LSS consistency relation is equivalent to extracting only the shift term, which is dominant in the squeezed limit, from the dark matter density fluctuations. Therefore, DHOST theories with the modified shift term violate the LSS consistency relation. In other words, the structure of DHOST theories resembles the structure of multicomponent fluids, and the Galilean transformation cannot eliminate the relative velocities of a scalar field and dark matter on large scales, thus violating the LSS consistency relation.

However, observables that trace the LSS consistency relation are not straightforwardly constructed. For example, \citet{Crisostomi:2019vhj} pointed out in Section V that taking the squeezed limit of the galaxy bispectrum does not directly test the violation of the LSS consistency relation in DHOST theories. The reason is that the bispectrum, which depends on the three wavenumbers $k_1$, $k_2$, and $k_3$, is symmetric with respect to these variables; taking the squeezed limit cancels out any change in the shift term that violates the consistency relation. Therefore, the authors proposed to measure the cross-bispectrum with other cosmic fluctuations, such as gravitational lensing effects, or to measure the trispectrum of galaxies, so that the effects of DHOST theories are not cancelled when the squeezed limit is taken.

Recent rapid developments in the analysis of galaxy three-point statistics, i.e. bispectra and three-point correlation functions (3PCFs), have allowed us to test the consistency relation. In principle, the three-point statistics (not restricted to the squeezed limit) are sensitive to the coefficient of the shift term of the galaxy density fluctuation. However, a critical problem remains to be solved: Most previous studies deal only with the isotropic, i.e. \emph{monopole}, component of the three-point statistics~\citep{Gil-Marin:2017MNRAS.465.1757G,Slepian:2016kfz,Pearson:2018MNRAS.478.4500P,Sugiyama:2018yzo,d'Amico2020JCAP...05..005D,Philcox:2022PhRvD.105d3517P,Cabass:2022arXiv220107238C,DAmico2022arXiv220111518D,Cabass:2022PhRvD.106d3506C}, and these analyses cannot efficiently constrain the coefficient of the nonlinear density field shift term. The reason is that in the monopole-only analysis, the coefficient of the density fluctuation shift term degenerates with the parameter $\sigma_8$, which represents the amplitude of the dark matter fluctuations~\citep{Sugiyama:2023arXiv230206808S}.

\citet{Yamauchi:2021arXiv210802382Y} pointed out that the use of nonlinear velocity fields in addition to nonlinear density fields is helpful in studying nonlinear gravitational effects. The nonlinear velocity field can be directly constrained by analysing the anisotropic, e.g. \emph{quadrupole}, component of the galaxy three-point statistics. Although the analysis of the anisotropic component of the galaxy three-point statistic is much less mature than the monopole-only analysis, some of us have successfully initiated such efforts. For example, \citet{Sugiyama:2018yzo} proposed a new basis for measuring the anisotropic bispectra and reported the significant detection of the anisotropic component from the galaxy catalogue from the Baryon Oscillation Spectroscopic Survey Data Release 12 (BOSS DR12)~\citep{Eisenstein:2011sa,Bolton:2012hz,Dawson:2012va,Alam:2015mbd}. \citet{Sugiyama:2020uil} analysed the anisotropic component of Baryon Acoustic Oscillations~\citep[BAOs;][]{Peebles:1970ag,Sunyaev:1970eu} using the anisotropic 2PCF and 3PCF on MultiDark-Patchy mock simulations~\citep[Patchy mock;][]{Klypin:2014kpa,Kitaura:2015uqa} that reproduce the BOSS DR12 galaxy data. \citet{D'Amico:2022arXiv220608327D} performed the first joint analysis of the monopole and quadrupole components of the power and bispectra measured from BOSS DR12 and constrained the standard cosmological parameters in the context of $\Lambda$CDM. \citet{Ivanov:2023arXiv230204414I} presented the results of an anisotropic bispectrum analysis including quadrupole and hexadecapole components measured from the BOSS DR12 data. In particular, \citet{Sugiyama:2023arXiv230206808S} applied the analysis method of \citet{Sugiyama:2020uil} to the BOSS galaxy data, based on the idea proposed by \citet{Yamauchi:2021arXiv210802382Y} to constrain the effects of gravitational nonlinearities arising from DHOST theories in a $\sigma_8$-independent manner.

The aim of this paper is to present, for the first time, an observational test of the LSS consistency relation in galaxy clustering in BOSS. We mostly follow the analysis method used in \citet{Sugiyama:2023arXiv230206808S} (hereafter referred to as S23). In order to ensure that the obtained results are applicable to as many different situations as possible, we propose a general parameterisation that includes modified gravity theories, multi-component fluids, and galaxy bias effects in the description of nonlinear density and velocity fields, thus constraining the LSS consistency relation in a broad framework. Conversely, when a violation of the LSS consistency relation is detected, a more specific model is required to provide a physical interpretation of the violation. We also present some specific examples of models that are and are not part of the parameterisation framework used in this paper. 

Our analysis uses a flat $\Lambda$CDM model as the fiducial cosmological model with the following parameters: matter density $\Omega_{\rm m0}=0.31$, Hubble constant $h\equiv H_0/(100\,{\rm km\, s^{-1}\, Mpc^{-1}})=0.676$, baryon density $\Omega_{\rm b0}h^2=0.022$, and spectral tilt $n_{\rm s}=0.97$, which are the same as those used in the final cosmological analysis in the BOSS project~\citep{Alam:2016hwk} and close to the best-fit values given by Planck2018~\citep{Aghanim:2018eyx}. In addition, we adopt a value for the total neutrino mass of $\sum m_{\nu}=0.06\, {\rm eV}$, which is close to the minimum allowed by neutrino oscillation experiments. We use the following publicly available libraries to perform theoretical calculations, measure 2PCF and 3PCF from galaxy data, and estimate parameter likelihoods using Markov Chain Monte Carlo (MCMC) methods: \textsc{Monte Python}~\citep{Brinckmann:2018cvx}, \textsc{CLASS}~\citep{Blas:2011rf}, \textsc{CUBA}~\citep{Hahn:2005CoPhC.168...78H}, \textsc{FFTW}~\citep{FFTW05}, and \textsc{FFTLog}~\citep{Hamilton:2000MNRAS.312..257H}.

The structure of this paper is as follows. Section~\ref{Sec:Equations} describes the theoretical model used in this paper and the parameters to be constrained; Section~\ref{Sec:Analysis} briefly summarises the data analysis methods; Section~\ref{Sec:Results} presents the results of the parameter constraints; Section~\ref{Sec:Conclusions} concludes the paper; Appendix~\ref{Sec:OtherParameters} summarises the results of other parameters not presented in the main text.

\section{Theoretical Background}
\label{Sec:Equations}

\subsection{Parameterisation of nonlinear fluctuations}
\label{Sec:Parameterization}

In this paper, we consider nonlinear effects up to the second order in perturbation theory since, in Section~\ref{Sec:Bispectrum}, we compute the 2PCF and 3PCF models based on the tree-level, taking into account nonlinear damping of the BAO. In Fourier space, the redshift-space density fluctuations of galaxies are expressed as~\citep[e.g,][]{Bernardeau:2001qr}
\begin{eqnarray}
    \delta_{1}(\VEC{k}) \hspace{-0.25cm}&=&\hspace{-0.25cm} 
    Z_1(\VEC{k}) \delta_{{\rm m}, 1}(\VEC{k}), \nonumber \\
    \delta_{2}(\VEC{k}) \hspace{-0.25cm}&=&\hspace{-0.25cm}  
    \int \frac{d^3p_1}{(2\pi)^3}
    \int \frac{d^3p_2}{(2\pi)^3}\delta_{\rm D}(\VEC{k}-\VEC{p}_{1}-\VEC{p}_2) \nonumber\\
     \hspace{-0.25cm} &\times& \hspace{-0.25cm}  
    Z_2(\VEC{p}_1,\VEC{p}_2) \delta_{{\rm m},1}(\VEC{p}_1)\delta_{{\rm m},1}(\VEC{p}_2),
    \label{Eq:D1D2}
\end{eqnarray}
where the numbers in the subscripts mean that the solution is of the first and second order in perturbation theory. The $\delta_{\rm m}$ appearing on the right-hand side represents the dark matter density fluctuation.

The first and second-order kernel functions $Z_1$ and $Z_2$ are given by~\citep{Kaiser:1987qv,Scoccimarro:1999ed}
\begin{eqnarray}
   \hspace{-0.45cm} Z_1(\VEC{k}) \hspace{-0.25cm}&=&\hspace{-0.25cm}   b_1 + f(\hat{k}\cdot\hat{n})^2, \nonumber \\
    \hspace{-0.45cm}Z_2(\VEC{k}_1,\VEC{k}_2)\hspace{-0.25cm}&=&\hspace{-0.25cm} F_2(\VEC{k}_1,\VEC{k}_2) + f (\hat{k}\cdot\hat{n})^2 G_2(\VEC{k}_1,\VEC{k}_2) \nonumber \\
    \hspace{-0.25cm}  &+& \hspace{-0.25cm}  \frac{f(\VEC{k}\cdot\hat{n})}{2} \left[ \frac{(\hat{k}_1\cdot\hat{n})}{k_1}Z_1(\VEC{k}_2) +  \frac{(\hat{k}_2\cdot\hat{n})}{k_2}Z_1(\VEC{k}_1)  \right],
    \label{Eq:Z1Z2}
\end{eqnarray}
where $b_1$ is the linear bias parameter, $f$ is the linear growth rate function, $\hat{n}$ is the unit vector that indicates the direction of the line of sight, and $\VEC{k}=\VEC{k}_1+\VEC{k}_2$. 

The kernel functions $F_2$ and $G_2$, which represent the second-order nonlinearity of the galaxy density fluctuation and the divergence of the galaxy velocity field, are decomposed into monopole, dipole, and quadrupole via the angle between $\VEC{k}_1$ and $\VEC{k}_2$ and are called the growth, shift, and tidal terms, respectively~\citep{Schmittfull:2015PhRvD..91d3530S}. We then introduce the following parameterisation for each coefficient of these terms:
\begin{eqnarray}
    F_2(\VEC{k}_1,\VEC{k}_2) \hspace{-0.25cm}&=&\hspace{-0.25cm} b_1 \left[F_{\rm g} + F_{\rm s}\, S(\VEC{k}_1,\VEC{k}_2) + F_{\rm t}\, T(\VEC{k}_1,\VEC{k}_2)  \right], \nonumber \\
    f G_2(\VEC{k}_1,\VEC{k}_2) \hspace{-0.25cm}&=&\hspace{-0.25cm} f \left[G_{\rm g} + G_{\rm s}\, S(\VEC{k}_1,\VEC{k}_2) + G_{\rm t}\, T(\VEC{k}_1,\VEC{k}_2)  \right],
    \label{Eq:F_2G_2}
\end{eqnarray}
where the subscripts 'g', 's', and 't' stand for 'growth', 'shift', and 'tidal', respectively. The scale-dependent functions characterising the shift and tidal terms are given by
\begin{eqnarray}
    S(\VEC{k}_1,\VEC{k}_2) \hspace{-0.25cm}&=& \hspace{-0.25cm} \frac{1}{2} (\hat{k}_1\cdot\hat{k}_2) \left( \frac{k_1}{k_2} + \frac{k_2}{k_1} \right), \nonumber \\
    T(\VEC{k}_1,\VEC{k}_2) \hspace{-0.25cm}&=& \hspace{-0.25cm}  (\hat{k}_1\cdot\hat{k}_2)^2 - \frac{1}{3}.
    \label{Eq:ST}
\end{eqnarray}

Since the linear and nonlinear fluctuations are proportional to $\sigma_8$ and $\sigma_8^2$, respectively, the parameters we focus on will appear in a degenerate form with $\sigma_8$, such as $(b_1\sigma_8)$, $(f\sigma_8)$, $(F_{\rm g}\sigma_8)$, $(F_{\rm s}\sigma_8)$, $(F_{\rm t}\sigma_8)$, $(G_{\rm g}\sigma_8)$, $(G_{\rm s}\sigma_8)$ and $(G_{\rm t}\sigma_8)$. Therefore, we introduce the following parameter to remove the dependence of $\sigma_8$ and express the violation of the LSS consistency relation~\citep{Yamauchi:2021arXiv210802382Y,Sugiyama:2023arXiv230206808S}:
\begin{eqnarray}
    E_{\rm s} \equiv \frac{(G_{\rm s}\sigma_8)}{(F_{\rm s}\sigma_8)}.
    \label{Eq:ES}
\end{eqnarray}
This $E_{\rm s}$ parameter satisfies $E_{\rm s}=1$ when $F_{\rm s}=G_{\rm s}$. For example, in GR, $E_{\rm s}=1$ since $F_{\rm s}=G_{\rm s}=1$. On the other hand, $E_{\rm s}\neq1$ is satisfied if either $F_{\rm s}$ or $G_{\rm s}$ or both are different from $1$, while keeping $F_{\rm s}\neq G_{\rm s}$. This means that the condition $E_{\rm s}\neq1$ indicates a violation of the LSS consistency relation. This is because taking the squeezed limit of the bispectrum corresponds to the operation of extracting only these shift terms. Note that $E_{\rm s}\neq1$ is a sufficient condition for proving the violation of the LSS consistency relation, not a necessary condition, since a particular theory may satisfy $F_{\rm s}=G_{\rm s}\neq1$.

\subsection{Consistency relation for the large-scale structure}
\label{Sec:ConsistencyRelation}

In this subsection, we show that the $E_{\rm s}$ parameter is useful for testing the violation of the LSS consistency relation. To do this, we focus on the bispectrum produced by the galaxy density fluctuations at three different redshifts, i.e.,
\begin{eqnarray}
   \hspace{-0.25cm}&& \hspace{-0.25cm}\langle \delta(\VEC{k}_1;z_1)\delta(\VEC{k}_2;z_2)\delta(\VEC{k}_3;z_3) \rangle \nonumber \\
   \hspace{-0.25cm}&=& \hspace{-0.25cm}  (2\pi)^3\delta_{\rm D}\left( \VEC{k}_1+\VEC{k}_2+\VEC{k}_3 \right)B(\VEC{k}_1,\VEC{k}_2;z_1,z_2,z_3),
   \label{Eq:Bz}
\end{eqnarray}
and take its squeezed limit $k_1\to0$. Note that this subsection is the only one in this paper that explicitly denotes the redshift dependence in the functions.

Satisfying the LSS consistency relation means that when taking the squeezed limit of the $n$-point statistics, the effect is described only by the contribution of the dark matter displacement vector evaluated at the origin. In real space, the linear displacement vector is $\mathbf{\Psi}_{\rm m,1}(\VEC{k}) = (i\VEC{k}/k^2)\delta_{\rm m,1}(\VEC{k})$, and in redshift space, it is computed by a linear transformation as 
\begin{eqnarray}
    \mathbf{\Psi}_{\rm s,1}(\VEC{k};z) = \MAT{R}(z)\cdot\mathbf{\Psi}_{\rm m,1}(\VEC{k};z),
\end{eqnarray}
where the transformation matrix $\MAT{R}$ is given by~\citep{Matsubara:2007wj}
\begin{eqnarray}
    \left[ \MAT{R}(z) \right]_{ij} =  \VEC{I}_{ij} + f(z)\,\hat{n}_i\hat{n}_j,
\end{eqnarray}
where $\MAT{I}$ is the three-dimensional identity matrix, and $i,j=1,2,3$.

For simplicity, we consider only the tree-level bispectrum. From Eq.~(\ref{Eq:D1D2}), the nonlinear contribution from a wavenumber $k_1$ or $k_2$ sufficiently smaller than the wavenumber $k$ of interest is given by the limit $k\gg k_1\to0$ or $k\gg k_2\to0$ and can be written as follows:
\begin{eqnarray}
    \delta_2(\VEC{k};z) \to 2\delta_{\rm m,1}(\VEC{k};z)
    \int \frac{d^3p}{(2\pi)} Z_2(\VEC{p},\VEC{k};z)|_{p\to0} \delta_{\rm m,1}(\VEC{p};z).
\end{eqnarray}
If $F_{\rm s}=G_{\rm s}=1$, then
\begin{eqnarray}
    \delta_2(\VEC{k};z) \to (-i\VEC{k}\cdot\overline{\mathbf{\Psi}}_{\rm s,1}(z)) \delta_{1}(\VEC{k};z),
\end{eqnarray}
where
\begin{eqnarray}
    \overline{\mathbf{\Psi}}_{\rm s,1}(z) = \int \frac{d^3p}{(2\pi)^3} \mathbf{\Psi}_{\rm s,1}(\VEC{p},z).
\end{eqnarray}
The above equation represents the inverse Fourier transform, and $\overline{\mathbf{\Psi}}_{\rm s,1}$ is the displacement vector at the origin $\VEC{x}=\VEC{0}$ and is independent of positions. In other words, $\overline{\mathbf{\Psi}}_{\rm s,1}$ can be interpreted as a large-scale flow of dark matter through the entire observation region.

Taking the squeezed limit $k_1\to0$ in the bispectrum of Eq.~(\ref{Eq:Bz}), $\delta(\VEC{k}_1;z_1)$ is only correlated with $\overline{\mathbf{\Psi}}_{\rm s,1}(z_2)$ or $\overline{\mathbf{\Psi}}_{\rm s,1}(z_3)$. Therefore,
\begin{eqnarray}
    \hspace{-0.25cm}&&\hspace{-0.25cm}\langle \delta(\VEC{k}_1;z_1)\delta(\VEC{k}_2;z_2)\delta(\VEC{k}_3;z_3) \rangle \nonumber \\
    \hspace{-0.25cm}&\underset{k_1\to0}{\to}&\hspace{-0.25cm} 
    \left\langle \left(-i\VEC{k}_2\cdot
        \left( \overline{\mathbf{\Psi}}_{\rm s,1}(z_2) - \overline{\mathbf{\Psi}}_{\rm s,1}(z_3) \right)\right) \delta_1(\VEC{k}_1;z_1) \right\rangle \nonumber \\
    \hspace{-0.25cm}&\times&\hspace{-0.25cm} 
    \langle \delta_1(\VEC{k}_2;z_2)\delta_1(\VEC{k}_3;z_3) \rangle,
\end{eqnarray}
leading to the LSS consistency relation
\begin{eqnarray}
    \hspace{-0.55cm}&&\hspace{-0.25cm} B(\VEC{k}_1,\VEC{k}_2;z_1,z_2,z_3) \nonumber \\
    \hspace{-0.55cm}&\underset{k_1\to0}{\to}&\hspace{-0.25cm}
    \tilde{Z}_1(\VEC{k}_1;z_1)
    \tilde{Z}_1(\VEC{k}_2;z_2)
    \tilde{Z}_1(\VEC{k}_2;z_3)
    \tilde{P}_{\rm lin}(k_1)
    \tilde{P}_{\rm lin}(k_2) \nonumber \\
    \hspace{-0.55cm}&\times&\hspace{-0.25cm} \left\{  \sigma_8(z_3) \frac{\VEC{k}_1\cdot\MAT{R}(z_3)\cdot\VEC{k}_2}{k_1^2}
    -\sigma_8(z_2) \frac{\VEC{k}_1\cdot \MAT{R}(z_2)\cdot\VEC{k}_2}{k_1^2}\right\},
    \label{Eq:CR}
\end{eqnarray}
where 
\begin{eqnarray}
    \langle \delta_{{\rm m},1}(\VEC{k};z)\delta_{{\rm m},1}(\VEC{k}';z)\rangle = (2\pi)^3\delta_{\rm D}(\VEC{k}+\VEC{k}')P_{\rm lin}(k;z)
\end{eqnarray}
gives the linear matter power spectrum, and $\tilde{P}_{\rm lin}$ and $\tilde{Z}_1$ are defined as $\tilde{P}_{\rm lin}(k) = P_{\rm lin}(k;z)/\sigma_8^2(z)$ and $\tilde{Z}_1(\VEC{k};z) = Z_1(\VEC{k};z)\sigma_8(z)$, respectively.

On the other hand, if $F_{\rm s}\neq1$ and $G_{\rm s}\neq 1$ are allowed, the tree-level bispectrum in the squeezed limit can be calculated as
\begin{eqnarray}
&& B(\VEC{k}_1,\VEC{k}_2;z_1,z_2,z_3) \nonumber \\
\hspace{-0.25cm}&\underset{k_1\to0}{\to}& \hspace{-0.25cm}
\tilde{Z}_1(\VEC{k}_1;z_1)\tilde{Z}_1(\VEC{k}_2;z_2) \tilde{Z_1}(\VEC{k}_2;z_3)
\tilde{P}_{\rm lin}(k_1)\tilde{P}_{\rm lin}(k_2) \nonumber \\
    \hspace{-0.25cm}&\times& \hspace{-0.25cm}
    \Bigg\{     
    \left( \frac{\VEC{k}_1\cdot\VEC{k}_2}{k_1^2} \right)
    \left[ (F_{\rm s}\sigma_8)(z_3) - (F_{\rm s}\sigma_8)(z_2) \right] \nonumber \\
   \hspace{-0.25cm} &+&\hspace{-0.25cm}
      \left( \frac{(\VEC{k}_1\cdot\hat{n})(\VEC{k}_2\cdot\hat{n})}{k_1^2} \right)
      \left[ (f\sigma_8)(z_3) - (f\sigma_8)(z_2)\right]
  \Bigg\}
     \nonumber \\
    \hspace{-0.25cm}&+&\hspace{-0.25cm}
    \left( \hat{k}_2\cdot\hat{n} \right)^2 \tilde{Z}_1(\VEC{k}_1;z_1)
    \left( \frac{\VEC{k}_1\cdot\VEC{k}_2}{k_1^2} \right) 
    \tilde{P}_{\rm lin}(k_1)\tilde{P}_{\rm lin}(k_2) 
    \nonumber \\
    \hspace{-0.25cm}&\times&\hspace{-0.25cm}
    \Bigg\{ (f\sigma_8)(z_3)(F_{\rm s}\sigma_8)(z_3)
    \left( E_{\rm s}(z_3)-1 \right) \tilde{Z}_1(\VEC{k}_2;z_2) \nonumber \\
     \hspace{-0.25cm}&-&\hspace{-0.25cm}
     (f\sigma_8)(z_2)(F_{\rm s}\sigma_8)(z_2)
    \left( E_{\rm s}(z_2)-1 \right)
\tilde{Z}_1(\VEC{k}_2;z_3) \Bigg\}.
    \label{Eq:CRFE}
\end{eqnarray}
Substituting $F_{\rm s}=E_{\rm s}=1$ into the above equation gives Eq.~(\ref{Eq:CR}). In other words, the LSS consistency relation is broken when $F_{\rm s}\neq1$ or $E_{\rm s}\neq1$. Also, as expected, $F_{\rm s}$ only appears in the form $(F_{\rm s}\sigma_8)$, indicating that the $E_{\rm s}$ parameter, which does not depend on $\sigma_8$, is the most appropriate for investigating the breakdown of the LSS consistency relation. However, since $\sigma_8>0$ by definition, $(F_{\rm s}\sigma_8<0)$ also implies the breakdown of the LSS consistency relation.

In actual observations, measuring correlators between galaxy density fluctuations at different redshifts is challenging. This is because the galaxy density fields at different redshifts are so far apart in the radial direction that they cannot be correlated. Therefore, it is common to measure the correlators of the galaxy density fields at equal time, so that $z_1=z_2=z_3$. In this case, the right-hand side of Eq.~(\ref{Eq:CRFE}) is always zero, because the remaining $k_2$ and $k_3$ dependencies are exchangeable when the squeezed limit $k_1\to0$ is taken between $k_1$, $k_2$ and $k_3$, on which the bispectrum is symmetrically dependent. This cancellation occurs even if $F_{\rm s}$ and $G_{\rm s}$ are scale-dependent functions with exchange symmetry (see Section~\ref{Sec:RelativeVelocities} for an example). Therefore, we propose to constrain $(F_{\rm s}\sigma_8)$ and $E_{\rm s}$ directly from the equal-time bispectrum (or 3PCF) without taking the squeezed limit. In this case, we simultaneously vary $(F_{\rm g}\sigma_8)$, $(F_{\rm t}\sigma_8)$, $(G_{\rm g}\sigma_8)$ and $(G_{\rm t}\sigma_8)$ as free parameters so that the results are valid in as general a situation as possible. More details on the physical meaning of these parameters are given in the following subsection.

We will focus only on the results for $E_{\rm s}$ in the main text, but the constraint results for the other nonlinear parameters, i.e. $(F_{\rm g}\sigma_8)$, $(F_{\rm s}\sigma_8)$, $(F_{\rm t}\sigma_8)$, $(G_{\rm g}\sigma_8)$, $(G_{\rm s}\sigma_8)$ and $(G_{\rm t}\sigma_8)$, are summarised in Appendix~\ref{Sec:OtherParameters}.

\subsection{Specific examples}
\label{Sec:Examples}

In this subsection, we discuss specific examples of models that are and are not covered by the parameterisation introduced in Eq.~(\ref{Eq:F_2G_2}).

\subsubsection{$\Lambda$CDM}
\label{Sec:LCDM}

In the $\Lambda$CDM model assuming $f^2=\Omega_{\rm m}$, the dark matter density and velocity fluctuations give~\citep[e.g.,][]{{Bernardeau:2001qr}}
\begin{eqnarray}
    F_{\rm g} \hspace{-0.25cm}&=&\hspace{-0.25cm} \frac{17}{21}, \quad
    F_{\rm s} = 1, \quad
    F_{\rm t} = \frac{2}{7}, \nonumber \\
    G_{\rm g} \hspace{-0.25cm}&=&\hspace{-0.25cm} \frac{13}{21}, \quad
    G_{\rm s} = 1, \quad
    G_{\rm t} = \frac{4}{7}.
    \label{Eq:Param_FG}
\end{eqnarray}
More generally, $f\sim \Omega_{\rm m}^{6/11}$ is a better approximation in a $\Lambda$CDM model. Then, the coefficients of the growth and tidal terms are also time-dependent~\citep[e.g.,][]{Fasiello:2022arXiv220510026F}. In particular, the following approximate formulae are given in the case where $f=\Omega_{\rm m}^{6/11}$~\citep{Bouchet:1992xz,Bouchet:1995A&A...296..575B,Yamauchi:2017ibz,Yamauchi:2021arXiv210802382Y}:
\begin{eqnarray}
    F_{\rm g} \hspace{-0.25cm}&=&\hspace{-0.25cm} F_{\rm s}-\frac{2}{3}F_{\rm t}, \quad
    F_{\rm s} = 1, \quad
    F_{\rm t} = \frac{2}{7}\, \Omega_{\rm m}^{3/572}, \nonumber \\
    G_{\rm g} \hspace{-0.25cm}&=&\hspace{-0.25cm} G_{\rm s}-\frac{2}{3}G_{\rm t}, \quad
    G_{\rm s} = 1, \quad
    G_{\rm t} = \frac{4}{7}\, \Omega_{\rm m}^{15/1144}.
    \label{Eq:Param_FG_ver2}
\end{eqnarray}
Note that the growth terms are not independent but are given by the shift and tidal terms. The reason is that the second-order kernel functions satisfy $F_2(\VEC{k},-\VEC{k})=0$ and $G_2(\VEC{k},-\VEC{k})=0$. This condition corresponds to the second-order density fluctuation smoothly approaching zero on large scales, i.e., $\delta_2(\VEC{k}\to0)\to0$, which represents the natural behaviour as a nonlinear effect.

\subsubsection{Horndeski theories}
\label{Sec:Horndeski}

Horndeski gravity theories are the most general scalar-tensor theory with second-order equations of motion for metric tensor and scalar fields~\citep{Horndeski:1974wa,Deffayet:2011gz,Kobayashi:2011nu}. In the Horndeski family of theories, the dark matter second-order density and velocity fields have time-dependent tidal terms, which are found to have a different time evolution than in the $\Lambda$CDM case~\citep[e.g.,][]{Takushima:2013foa}:
\begin{eqnarray}
    F_{\rm g} \hspace{-0.25cm}&=&\hspace{-0.25cm} F_{\rm s}-\frac{2}{3}F_{\rm t}, \quad
    F_{\rm s} = 1, \quad
    F_{\rm t} = \frac{2}{7}\, \lambda_{\delta}, \nonumber \\
    G_{\rm g} \hspace{-0.25cm}&=&\hspace{-0.25cm} G_{\rm s}-\frac{2}{3}G_{\rm t}, \quad
    G_{\rm s} = 1, \quad
    G_{\rm t} = \frac{4}{7}\, \lambda_{\theta},
\end{eqnarray}
where $\lambda_{\delta}$ and $\lambda_{\theta}$ are time-dependent functions, and they are related to each other as
\begin{eqnarray}
    \lambda_{\theta} = \lambda_{\delta}\left[ 1 + \frac{1}{2f}\frac{d \ln \lambda_{\delta}}{d \ln a} \right]
\end{eqnarray}
with $a$ being the scale factor.

Compared to Eq.~(\ref{Eq:Param_FG_ver2}), \citet{Yamauchi:2017ibz} proposed to test the nonlinearity of Horndeski theories by using the parameterisation 
\begin{eqnarray}
    \lambda_{\delta} = \Omega_{\rm m}^{\xi}.
\end{eqnarray}
While it is widely used in linear theory to test modified gravity theories by constraining $\gamma=\log_{\Omega_{\rm m}}(f)$, this parameterisation is an extension to nonlinear effects. The authors also showed that $\xi$ contains new information compared to $\gamma$ in the test of Horndeski theories by giving a specific model that satisfies $\gamma=6/11$ and $\xi\neq 3/573$.

\subsubsection{DHOST theories}
\label{Sec:DHOST}

Going beyond Horndeski theories, DHOST theories have been recently discovered~\citep[for reviews,][]{Langlois:2018dxi,Kobayashi:2019hrl}. Even though DHOST theories have higher-order equations of motion, they reduce in the end to a second-order system thanks to the degeneracy between the kinetic terms of the scalar and metric fields, leading to healthy scalar-tensor theories. In the dark matter second-order density and velocity fields in these theories, in addition to the tidal terms, the shift terms also become time-dependent and deviate from the $\Lambda$CDM prediction of $1$~\citep[e.g.,][]{Hirano:2018uar}:
\begin{eqnarray}
    F_{\rm g} \hspace{-0.25cm}&=&\hspace{-0.25cm} F_{\rm s}-\frac{2}{3}F_{\rm t}, \quad
    F_{\rm s} = \kappa_{\delta}, \quad
    F_{\rm t} = \frac{2}{7}\, \lambda_{\delta}, \nonumber \\
    G_{\rm g} \hspace{-0.25cm}&=&\hspace{-0.25cm} G_{\rm s}-\frac{2}{3}G_{\rm t}, \quad
    G_{\rm s} = \kappa_{\theta}, \quad
    G_{\rm t} = \frac{4}{7}\, \lambda_{\theta},
\end{eqnarray}
where $\kappa_{\delta}$ and $\kappa_{\theta}$ are time-dependent functions, and they are related via 
\begin{eqnarray}
    \kappa_{\theta} = 2\kappa_{\delta} \left[ 1 + \frac{1}{2f}\frac{d \ln \kappa_{\delta}}{d \ln a}  \right]- 1.
\end{eqnarray}

\citet{Yamauchi:2021arXiv210802382Y} proposed the following parameterisation for observationally testing DHOST theories,
\begin{eqnarray}
    \frac{f}{\kappa_{\delta}} = \Omega_{\rm m}^{\xi_f}, \quad
    \frac{\kappa_{\theta}}{\kappa_{\delta}} = \Omega_{\rm m}^{\xi_{\rm s}}, \quad
    \frac{\lambda_{\theta}}{\kappa_{\delta}} = \Omega_{\rm m}^{\xi_{\rm t}},
    \label{Eq:OmegaXI}
\end{eqnarray}
and pointed out that any non-vanishing value of $\xi_{\rm s}$ can be treated as a clear signal of the existence of a gravity theory beyond Horndeski theories. S23 constrained these index parameters using the BOSS DR12 galaxies and the results are
\begin{eqnarray}
    -0.907 \hspace{-0.25cm}&<&\hspace{-0.25cm} \xi_f < 2.447, \nonumber \\
    -1.655 \hspace{-0.25cm}&<&\hspace{-0.25cm} \xi_{\rm t}, \nonumber \\
    -0.504 \hspace{-0.25cm}&<&\hspace{-0.25cm} \xi_{\rm s}
\end{eqnarray}
at the $95\%$ confidence level. Note that the upper bounds on $\xi_{\rm t}$ and $\xi_{\rm s}$ are not given because $(\lambda_{\theta}/\kappa_{\delta})$ and $(\kappa_{\theta}/\kappa_{\delta})$ are consistent with zero within the $95\%$ error, and $\xi_{\rm t}$ and $\xi_{\rm s}$ can each take infinitely large values as $(\lambda_{\theta}/\kappa_{\delta})$ and $(\kappa_{\theta}/\kappa_{\delta})$ approach zero.

The middle equation in Eq.~(\ref{Eq:OmegaXI}) corresponds to the $E_{\rm s}$ parameter introduced in Eq.~(\ref{Eq:ES}), i.e., $E_{\rm s}=\kappa_{\theta}/\kappa_{\delta}$. However, while this paper follows the analysis approach of S23, it is no longer restricted to DHOST theories and assumes a more general situation that includes effects other than those of modified gravity theories. Even if the results of this paper are used to constrain DHOST theories, there are two obvious differences with S23. First, while S23 assumes the standard bias theory given in Section~\ref{Sec:StandardBias}, this paper assumes the existence of more general bias parameters in Section~\ref{Sec:ExtendedBias} and varies all nonlinear parameters $F_{\rm g}$, $F_{\rm s}$, $F_{\rm t}$, $G_{\rm g}$, $G_{\rm s}$, and $G_{\rm t}$ as free parameters. Second, the parameterisation given in Eq.~(\ref{Eq:OmegaXI}) implicitly assumes $E_{\rm s}>0$, whereas this paper allows negative $E_{\rm s}$.

\subsubsection{5D brane-world model}
\label{Sec:BraneWorld}

The normal branch of the 5-dimensional Dvali-Gabadadze-Porrati brane-world model~\citep[nDGP;][]{Dvali:2000hr}, which is a kind of modified gravity theories with extra dimensions, has been well studied. However, since the effects of the extra dimension can be described effectively as a scalar field, this brane-world model can be subsumed into scalar-tensor theories.

The nDGP model is characterised by a nonlinear function that modifies the Poisson equation~\citep[e.g.,][]{Koyama:2009me,Bose:2016qun},
\begin{eqnarray}
    \gamma_2(\VEC{k}_1,\VEC{k}_2) \propto \left( 1 - \left( \hat{k}_1\cdot\hat{k}_2 \right)^2 \right)
    = \frac{2}{3} - T\left( \VEC{k}_1,\VEC{k}_2 \right),
\end{eqnarray}
varying the tidal term from GR. Once the tidal term is determined, the growth term is also determined by $\gamma_2(\VEC{k},-\VEC{k})=0$. Thus, the nDGP model can be described by a parameterisation similar to the Horndeski theories in Section~\ref{Sec:Horndeski}.

\subsubsection{$f(R)$ gravity}

In this subsection, we discuss the Hu-Sawicki model~\citep{Hu:2007nk} of $f(R)$ gravity~\citep[see][for reviews]{Capozziello:2007ec,Sotiriou:2008rp}, which is widely used in cosmology. The Hu-Sawicki model predicts nonlinear functions that modify the Poisson equation as follows~\citep[e.g,][]{Koyama:2009me,Bose:2016qun}:
\begin{eqnarray}
    \gamma_2(\VEC{k}_1,\VEC{k}_2) \hspace{-0.25cm}&\propto&\hspace{-0.25cm} \left( \frac{k_{12}}{aH} \right)^2\frac{1}{\Pi(k_{12})\Pi(k_{1})\Pi(k_{2})}, \nonumber\\
    \Pi(k) \hspace{-0.25cm} &=& \hspace{-0.25cm} \left( \frac{k}{a} \right)^2 + \frac{H_0^2\left( \Omega_{\rm m0}-4a^3\left( \Omega_{\rm m0}-1 \right) \right)^3}{2|f_{R0}|a^9\left( 3\Omega_{\rm m0}-4 \right)^2},
    \label{Eq:fR}
\end{eqnarray}
where $\VEC{k}_{12}=\VEC{k}_1+\VEC{k}_2$, $H_0$ and $\Omega_{\rm m0}$ are the Hubble parameter and the current matter density fraction, respectively, and $|f_{R0}|$ is a free parameter of the theory. Note that in this model, unlike the other models presented in this paper, even linear density fluctuations cannot separate the time dependence from the wavenumber dependence.

The relationship between $f(R)$ gravity and scalar-tensor gravity, and their possible equivalence, has been extensively studied~\citep[e.g.,][]{Sotiriou:2006CQGra..23.5117S}. In particular, the gravitational nonlinear effects that are the focus of this paper have been discussed in relation to Horndeski theories in Appendix B of \citet{Bose:2016qun}. However, the gravitational nonlinearities obtained from the Hu-Sawicki model differ from those of the Horndeski type discussed in Section~\ref{Sec:Horndeski}. We suspect that this difference is due to the fact that the nonlinear effects given in Section~\ref{Sec:Horndeski} focus only on the terms for which the spatial derivative is most active in the quasi-static approximation, and neglect the terms corresponding to the mass terms of the scalar field, while the $f(R)$ gravity model retains such terms. Since a detailed proof of this is beyond the scope of this paper, we limit ourselves to pointing out that the nonlinear effect of the Hu-Sawicki model given by Eq.~(\ref{Eq:fR}) does not fit into the parameterisation framework used in this paper.

\subsubsection{Nearly horizon scales}
\label{Sec:Horizon}

This paper focuses on the LSS consistency relation in the sub-horizon limit. Thus, even in the GR case, at large scales close to the horizon scale, there are additional correction terms for the nonlinear effects given in Section~\ref{Sec:LCDM}, which are derived in the Newtonian limit. For example, following~\citet{Tram:2016JCAP...05..058T}, a correction term proportional to $(aH/k)^2$ arises for $F_{\rm g}$, $F_{\rm s}$ and $F_{\rm t}$, respectively, and a new scale dependence emerges as follows:
\begin{eqnarray}
    {\cal K}(\VEC{k}_1,\VEC{k}_2) \propto \left( \frac{aH}{k_{12}} \right)^2 \left( \frac{k_1}{k_2} - \frac{k_2}{k_1} \right)^2.
\end{eqnarray}
Therefore, the parameterisation used in this paper is only valid at the sub-horizon scale. See also \citet{Creminelli:2013mca} for a fully relativistic consistency relation. \citet{Inomata:2023arXiv230410559I} also provide a detailed study of squeezed $n$-point functions in synchronous gauge.

\subsubsection{Massive neutrinos}
\label{Sec:MassiveNeutrinos}

In the remainder of this section, we denote the nonlinear parameters for dark matter in a gravity theory described in Sections~\ref{Sec:LCDM}-\ref{Sec:BraneWorld} as $F_{\rm g,s,t}^{(\rm m)}$ and $G_{\rm g,s,t}^{(\rm m)}$ and the additional correction terms for them as $\Delta F_{\rm g,s,t}^{(\rm m)}$ and $\Delta G_{\rm g,s,t}^{(\rm m)}$: i.e.,
\begin{eqnarray}
    F_{\rm g,s,t} \hspace{-0.25cm}&=&\hspace{-0.25cm} F_{\rm g,s,t}^{(\rm m)} + \Delta F_{\rm g,s,t}, \nonumber \\
    G_{\rm g,s,t} \hspace{-0.25cm}&=&\hspace{-0.25cm} G_{\rm g,s,t}^{(\rm m)} + \Delta G_{\rm g,s,t}.
\end{eqnarray}

Massive neutrinos can modify the second-order kernel functions, in which case the following correction terms are added to the nonlinear parameters~\citep{Kamalinejad:2020arXiv201100899K}:
\begin{eqnarray}
    \hspace{-0.7cm}\Delta F_{\rm g} \hspace{-0.25cm}&=&\hspace{-0.25cm} \frac{4}{245}f_{\nu}, \quad
    \Delta F_{\rm s} = 0, \quad
    \Delta F_{\rm t} =  - \frac{6}{245}f_{\nu}, \nonumber \\
    \hspace{-0.7cm}\Delta G_{\rm g} \hspace{-0.25cm}&=&\hspace{-0.25cm}  - \frac{83}{245}f_{\nu}, \quad
    \Delta G_{\rm s} =  - \frac{3}{5}f_{\nu}, \quad
    \Delta G_{\rm t} =  - \frac{96}{245}f_{\nu},
    \label{Eq:Neutrino}
\end{eqnarray}
where the neutrino density fraction $f_{\nu}$ is given by
\begin{eqnarray}
    f_{\nu} = \frac{\Omega_{\nu}}{\Omega_{\rm m}}
\end{eqnarray}
with $\Omega_{\nu}$ and $\Omega_{\rm m}$ being the neutrino and matter energy densities in units of the critical density, respectively. It is important to note that the massive neutrinos do not change the shift term of the density fluctuation but correct the shift term of the velocity fluctuation. 

Finally, we estimate the extent to which $E_{\rm s}$ deviates from $1$ in the presence of massive neutrinos. The neutrino density fraction is given by~\citep[e.g.,][]{Takada:2006PhRvD..73h3520T}
\begin{eqnarray}
    f_{\nu} = 0.05 \left( \frac{\sum m_{\nu}}{0.658\, {\rm eV}} \right) \left( \frac{0.14}{\Omega_{\rm m}h^2} \right).
    \label{Eq:fnu}
\end{eqnarray}
According to current observations, the upper limit for the total neutrino mass is $\sum m_{\nu}\lesssim 0.1\, {\rm eV}$ at $95\%$ CL~\citep[e.g.,][]{DiValentino2021PhRvD.104h3504D}. Consequently, substituting $\sum m_{\nu}= 0.1\, {\rm eV}$ into Eq.~(\ref{Eq:fnu}), the expected value of $E_{\rm s}-1=-(3/5) f_{\nu}$ is then
\begin{eqnarray}
    E_{\rm s}-1 = - 0.0046\, \left( \frac{\sum m_{\nu}}{0.1\, {\rm eV}} \right)\left( \frac{0.14}{\Omega_{\rm m}h^2} \right).
\end{eqnarray}
Thus, the impact of the neutrino masses in $E_{\rm s}$ would be minimal, since the $1\mathchar`-\sigma$ error for $E_{\rm s}$ obtained from the current BOSS data is about $3$ in Section~\ref{Sec:Es}. Put differently, it would be challenging to strongly constrain neutrino masses in the future using only $E_{\rm s}$.

\subsubsection{Standard bias effects}
\label{Sec:StandardBias}

In standard bias theory, the nonlinear bias parameters connecting the galaxy density field and the dark matter density field appear in the growth and tidal terms of the density fluctuations. Thus, they are added to $F_{\rm g}$ and $F_{\rm t}$ as follows~\citep[for a review, see][]{Desjacques:2016bnm}:
\begin{eqnarray}
    \Delta F_{\rm g} \hspace{-0.25cm}&=&\hspace{-0.25cm} \frac{1}{2}\frac{b_2}{b_1}, \quad
    \Delta F_{\rm t} = \frac{b_{\rm t}}{b_1},
\end{eqnarray}
where $b_2$ and $b_{\rm t}$ denote the local nonlinear bias parameter and the tidal bias parameter, respectively. In this case, the condition $F_2(\VEC{k},-\VEC{k})=0$, which is satisfied in the absence of the bias effect, does not hold, and $F_{\rm g}$ should be treated as an independent parameter, while $G_{\rm g}$ remains dependent.

\subsubsection{Relative velocities}
\label{Sec:RelativeVelocities}

The relative velocity effects of baryons and cold dark matter, together with a corresponding bias parameter, enter the galaxy density fluctuation with a quadratic form~\citep{Dalal:2010yt}. The resulting shift term is modified in the second-order density fluctuation~\citep{Yoo:2011tq}:
\begin{eqnarray}
   \Delta F_{\rm s} = - \frac{b_r}{b_1} \frac{T_{\rm rv}(k_1)T_{\rm rv}(k_2)}{T_{\rm m}(k_1)T_{\rm m}(k_2)},
\end{eqnarray}
where $b_r$ denotes the relative velocity bias parameter, $T_{\rm rv}$ is the relative velocity transfer function, and $T_{\rm m}$ is the dark matter transfer function. This relative velocity effect on galaxy clustering has been measured using galaxy power spectra and 3PCFs, but its signature has not yet been detected~\citep{Yoo:2013qla,Beutler:2015tla,Slepian:2016nfb}.

The relative velocity effect is obtained by the ratio of the relative velocity to the dark matter transfer functions $T_{\rm rv}(k)/T_{\rm m}(k)$, which is scale-dependent and therefore does not fit into the parameterisation framework of this paper. However, if a signal with $E_{\rm s}\neq1$ is detected, a correct physical interpretation would require a reanalysis to account for this possible relative velocity effect.

\subsubsection{Extended bias effects}
\label{Sec:ExtendedBias}

In this paper, we discuss the possibility of extended bias theories. For example, in specific gravity theories, such as DHOST theories, the coefficient of the density fluctuation shift term $F_{\rm s}$ deviates from $1$, violating the LSS consistency relation. On the other hand, \citet{Fujita:2020JCAP...10..059F} showed that the standard bias theory is reproduced in theories that satisfy the LSS consistency relation. In other words, for DHOST theories with $F_{\rm s}\neq1$, there may be an additional bias effect in $F_{\rm s}$. Since the shift term is described by the product of the displacement vector and the density fluctuation, the bias of the shift term may be related to the bias effect of the displacement vector. Furthermore, since the time derivative of the displacement vector is a velocity field, the bias effect of the displacement vector may induce the bias effect of linear and nonlinear velocity fields (see also Section $9.12$ in S23).

Based on the above considerations, we assume that bias effects occur for all nonlinear parameters:
\begin{eqnarray}
    \hspace{-0.7cm}\Delta F_{\rm g} \hspace{-0.25cm}&=&\hspace{-0.25cm} \frac{1}{2}\frac{b_2}{b_1}, \quad
    \Delta F_{\rm s} = \frac{b_{\rm s}}{b_1}, \quad
    \Delta F_{\rm t} = \frac{b_{\rm t}}{b_1}, \nonumber \\
    \hspace{-0.7cm}\Delta G_{\rm g} \hspace{-0.25cm}&=&\hspace{-0.25cm} \frac{1}{2}b_{v2}, \quad
    \Delta G_{\rm s} =  b_{v\rm s}, \quad
    \Delta G_{\rm t} =  b_{v\rm t},
\end{eqnarray}
where $b_{\rm s}$ is the shift bias in the second-order density fluctuation, and $b_{v2}$, $b_{v\rm s}$, and $b_{v\rm t}$ are the nonlinear local bias, shift bias, and tidal bias in the velocity fluctuation. In such an extended bias theory, the condition $G_2(\VEC{k},-\VEC{k})=0$ no longer holds, and $G_{\rm g}$ should also be treated as an independent parameter. Since the assumption of a linear velocity bias does not change the form of Eq.~(\ref{Eq:Z1Z2}), but only multiplies the velocity bias parameter $b_v$ by $f$, we implicitly assume a linear velocity bias and use $f$ as it is. Of course, numerical experiments such as $N$-body simulations of dark matter, including effects such as DHOST theories, are needed to verify this consideration. Such studies are left as future work.

\subsection{Bispectrum and 3PCF models}
\label{Sec:Bispectrum}

The leading order galaxy power spectrum and bispectrum in perturbation theory are given by
\begin{eqnarray}
    P(\VEC{k}) \hspace{-0.25cm}&=&\hspace{-0.25cm} [Z_1(\VEC{k})]^2 P_{\rm lin}(k), \nonumber \\
    B(\VEC{k}_1,\VEC{k}_2) \hspace{-0.25cm}&=&\hspace{-0.25cm} 2 Z_2(\VEC{k}_1,\VEC{k}_2) Z_1(\VEC{k}_1)Z_1(\VEC{k}_2) P_{\rm lin}(k_1)P_{\rm lin}(k_2) \nonumber \\
    \hspace{-0.25cm} &+&\hspace{-0.25cm} (\VEC{k}_1\leftrightarrow\VEC{k}_3) + (\VEC{k}_2\leftrightarrow\VEC{k}_3),
   \label{Eq:Tree_PB}
\end{eqnarray}
where $\VEC{k}_1+\VEC{k}_2+\VEC{k}_3=0$.

The theoretical models in Eq.~(\ref{Eq:Tree_PB}) work well in principle on large scales around and above $100\hMpc$, but they cannot describe the nonlinear decay of the signal of the Baryon Acoustic Oscillations~\citep[BAOs;][]{Peebles:1970ag,Sunyaev:1970eu} that occurs around $100\hMpc$. In order to include the effects of the nonlinear decay of the BAO, while preserving the form of the leading-order solutions of the power spectrum and the bispectrum, we use the following theoretical models, which are obtained by re-summing the IR modes appearing in the expansion via perturbation theory~\citep{Eisenstein:2006nj,Sugiyama:2020uil}:
\begin{eqnarray}
    P(\VEC{k})\hspace{-0.25cm} &=& \hspace{-0.25cm}\left[ Z_1(\VEC{k}) \right]^2 \left[{\cal D}^2(\VEC{k})P_{\rm w}(k) + P_{\rm nw}(k) \right],
    \nonumber \\
    B(\VEC{k}_1,\VEC{k}_2) 
    \hspace{-0.25cm}&=&\hspace{-0.25cm}  2\, Z_2(\VEC{k}_1,\VEC{k}_2)Z_1(\VEC{k}_1)Z_1(\VEC{k}_2) \nonumber \\
    \hspace{-0.25cm}&\times&\hspace{-0.25cm} \Big\{ {\cal D}(\VEC{k}_1){\cal D}(\VEC{k}_2){\cal D}(\VEC{k}_3) P_{\rm w}(k_1)P_{\rm w}(k_2) \nonumber \\
    \hspace{-0.25cm}&+&\hspace{-0.25cm} {\cal D}^2(\VEC{k}_1) P_{\rm w}(k_1)P_{\rm nw}(k_2) + {\cal D}^2(\VEC{k}_2) P_{\rm nw}(k_1)P_{\rm w}(k_2) \nonumber \\
    \hspace{-0.25cm}&+&\hspace{-0.25cm} P_{\rm nw}(k_1)P_{\rm nw}(k_2)\Big\} 
    +  (\VEC{k}_1\leftrightarrow\VEC{k}_3) + (\VEC{k}_2\leftrightarrow\VEC{k}_3),
    \label{Eq:PB_IR}
\end{eqnarray}
where $P_{\rm lin}$ is decomposed into two parts: the "no-wiggle (nw)" part $P_{\rm nw}$, which is a smooth version of $P_{\rm lin}$ with the baryon oscillations removed~\citep{Eisenstein:1997ik}, and the "wiggle (w)" part defined as $P_{\rm w}=P_{\rm lin}-P_{\rm nw}$. The nonlinear BAO degradation is represented by the two-dimensional Gaussian damping factor derived from a differential motion of Lagrangian displacements~\citep{Eisenstein:2006nj,Crocce:2007dt,Matsubara:2007wj}:
\begin{eqnarray}
    {\cal D}(\VEC{k}) = \exp\left( -\frac{ k^2(1-\mu^2)\sigma_{\perp}^2 + k^2\mu^2 \sigma^2_{\parallel} }{2} \right),
    \label{Eq:Damping}
\end{eqnarray}
where $\mu=\hat{k}\cdot\hat{n}$. We compute the radial and transverse components of the smoothing parameters, $\sigma_{\perp}$ and $\sigma_{\parallel}$, using the Zel'dovich approximation~\citep{Zeldovich:1970,Crocce:2007dt,Matsubara:2007wj}:
\begin{eqnarray}
    \sigma^2_{\perp} \hspace{-0.25cm}&=&\hspace{-0.25cm} 
    \frac{1}{3}\int \frac{dp}{2\pi^2} P_{\rm lin}(p), \nonumber \\
    \sigma^2_{\parallel} \hspace{-0.25cm}&=&\hspace{-0.25cm} (1+f)^2\, \sigma^2_{\perp}.
    \label{Eq:sigma_sigma}
\end{eqnarray}

We decompose the power spectrum into multipole components using Legendre polynomial functions ${\cal L}_{\ell}$~\citep[e.g.,][]{Hamilton:1997zq}:
\begin{eqnarray}
    P(\VEC{k}) = \sum_{\ell} P_{\ell}(k) {\cal L}_{\ell}(\hat{k}\cdot\hat{n}).
\end{eqnarray}
The multipole components of the power spectrum are then related to those of the 2PCF by a one-dimensional Hankel transformation:
\begin{eqnarray}
    \xi_{\ell}(r) 
    \hspace{-0.25cm}&=&\hspace{-0.25cm}
    i^{\ell} \int \frac{dkk^2}{2\pi^2} j_{\ell}(rk)P_{\ell}(k),
\end{eqnarray}
where $j_{\ell}$ is the $\ell$-th order spherical Bessel function. The multipole index $\ell$ refers to the expansion with respect to the line-of-sight dependence due to the Redshift Space Distortion effect~\citep[RSD;][]{Kaiser:1987qv} and the Alcock-Paczy\'{n}ski effect~\citep[AP;][]{Alcock:1979mp}. The components with $\ell=0$, $2$, and $4$ are called monopole, quadrupole, and hexadecapole, respectively; the components with $\ell>0$ are caused only by the RSD effect and the AP effect.

We adopt the decomposition formalism of the bispectrum into multipole components using tri-polar spherical harmonic (TripoSH) base functions~\citep{Sugiyama:2018yzo}:
\begin{eqnarray}
    B(\VEC{k}_1,\VEC{k}_2)=\hspace{-0.25cm} \sum_{\ell_1+\ell_2+\ell={\rm even}}\hspace{-0.25cm}  B_{\ell_1\ell_2\ell}(k_1,k_2)\,{\cal S}_{\ell_1\ell_2\ell}(\hat{k}_1,\hat{k}_2,\hat{n}),
\end{eqnarray}
where the TripoSH base functions are given by
\begin{eqnarray}
    {\cal S}_{\ell_1\ell_2\ell}(\hat{k}_1,\hat{k}_2,\hat{n}) 
	\hspace{-0.25cm}&=&\hspace{-0.25cm}
   \frac{4\pi}{h_{\ell_1\ell_2\ell}} 
   \sum_{m_1m_2m} 
   \left( \begin{smallmatrix} \ell_1 & \ell_2 & \ell \\ m_1 & m_2 & m \end{smallmatrix}  \right)\nonumber \\ 
    \hspace{-0.25cm}&\times&\hspace{-0.25cm}
    Y_{\ell_1 m_1}(\hat{k}_1) Y_{\ell_2 m_2}(\hat{k}_2) Y_{\ell m}(\hat{n}),
    \label{Eq:Slll}
\end{eqnarray} 
with
\begin{eqnarray}
    h_{\ell_1\ell_2\ell}= \sqrt{ \frac{(2\ell_1+1)(2\ell_2+1)(2\ell+1)}{4\pi}}
   \left( \begin{smallmatrix} \ell_1 & \ell_2 & \ell \\ 0 & 0 & 0 \end{smallmatrix}  \right).
\end{eqnarray}
The multipole components of the bispectrum are then related to those of the 3PCF by a two-dimensional Hankel transformation:
\begin{eqnarray}
	\zeta_{\ell_1\ell_2\ell}(r_1,r_2) 
    \hspace{-0.25cm}&=&\hspace{-0.25cm}
    i^{\ell_1+\ell_2} \int \frac{dk_1k_1^2}{2\pi^2} \int \frac{dk_2k_2^2}{2\pi^2} \nonumber \\
	\hspace{-0.25cm}&\times&\hspace{-0.25cm}
    j_{\ell_1}(r_1k_1) j_{\ell_2}(r_2k_2) B_{\ell_1\ell_2\ell}(k_1,k_2).
    \label{Eq:B_to_zeta}
\end{eqnarray}
The multipole index $\ell$ appearing in $B_{\ell_1\ell_2\ell}$ or $\zeta_{\ell_1\ell_2\ell}$ is associated with the multipole expansion w.r.t. the line of sight, just as the index $\ell$ in the power spectrum, $P_{\ell}$. 
%the same meaning as that appearing in $P_{\ell}$ or $\xi_{\ell}$.

\subsection{Theoretical predictions}

\begin{figure}
    \centering
    \scalebox{0.92}{\includegraphics[width=\columnwidth]{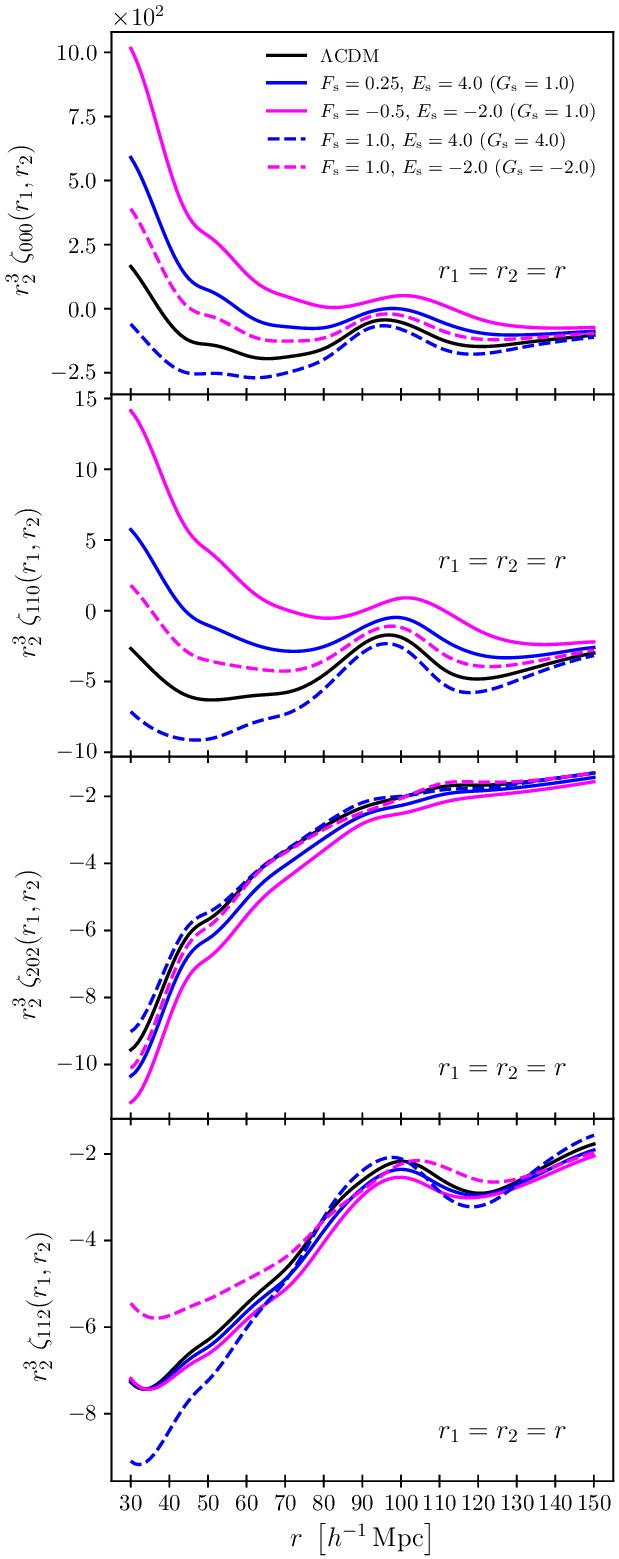}}
    \caption{Multipole components of the 3PCF, i.e. $\zeta_{000}$, $\zeta_{110}$, $\zeta_{202}$ and $\zeta_{112}$, at $z=0.61$, calculated from the theoretical model in Eq.~(\ref{Eq:PB_IR}), when the coefficients of the shift terms of the density or velocity fluctuations, i.e. $F_{\rm s}$ or $G_{\rm s}$, vary from $1$. The results are shown for $F_{\rm s}=G_{\rm s}=1$ (black solid), $F_{\rm s}=0.25$ and $G_{\rm s}=1$ (blue solid), $F_{\rm s}=-0.5$ and $G_{\rm s}=1$ (magenta solid), $F_{\rm s}=1$ and $G_{\rm s}=4.0$ (blue dashed), $F_{\rm s}=1$ and $G_{\rm s}=-2.0$ (magenta dashed). For the sake of simplicity, the plot is made as a function of $r_1=r_2=r$. The cosmological parameters used to draw this plot are given in Section~\ref{Sec:Introduction}, and the assumed linear bias is $b_1=2$, and the assumed nonlinear biases are zero, i.e. $b_2=b_{\rm t}=0$.
}
    \label{fig:3PCF_model}
\end{figure}

Figure~\ref{fig:3PCF_model} shows how the multipole components of the 3PCF are affected when the coefficients of the shift terms for density and velocity fluctuations, i.e. $F_{\rm s}$ and $G_{\rm s}$, are changed from $1$. Since Section~\ref{Sec:Es} will show that the 1-$\sigma$ error for $E_{\rm s}$ is about $3$, we add $\pm3$ to the $E_{\rm s}=1$ value in $\Lambda$CDM to compute the cases where $E_{\rm s} = -2$ and $E_{\rm s}=4$. In other words, we compute the four cases for $(F_{\rm s}=0.25, G_{\rm s}=1.0)$, $(F_{\rm s}=-0.5, G_{\rm s}=1.0)$, $(F_{\rm s}=1.0, G_{\rm s}=4.0)$, and $(F_{\rm s}=1.0, G_{\rm s}=-2.0)$.

Focusing on the monopole components, i.e. $\zeta_{000}$ and $\zeta_{110}$, the effect of changing the value of $F_{\rm s}$ is more significant than when $G_{\rm s}$ is changed. This result suggests that the monopole component can constrain $F_{\rm s}$ well. Next, we look at the quadrupole components, i.e. $\zeta_{202}$ and $\zeta_{112}$. Again, the change in $F_{\rm s}$ can affect them more than in $G_{\rm s}$, but the difference is less than for the monopole component. This fact means that $G_{\rm s}$ or $E_{\rm s}$ is determined in the quadrupole component after $F_{\rm s}$ has been determined in the monopole component.

Of course, in the actual MCMC analysis, not only $F_{\rm s}$ and $G_{\rm s}$ are varied, but also $F_{\rm g}$, $F_{\rm t}$, $G_{\rm g}$ and $G_{\rm t}$. The influence of all these parameters on the 3PCF multipole can be seen in Figures $1$-$2$ of S23.

\section{Data Analysis Methodology}
\label{Sec:Analysis}

Our data analysis methods are summarised below. See S23 for details.

\begin{enumerate}

    \item To simplify the correction for window function effects, the 2PCF and the 3PCF are used instead of the power spectrum and the bispectrum in Fourier space, following~\citep{Sugiyama:2020uil,Sugiyama:2023arXiv230206808S}.

        \vspace{2mm}

    \item Only large scales in the range $80\hMpc\leq r \leq150\hMpc$ are used, where the 2PCF and 3PCF models~(\ref{Eq:PB_IR}) are expected to work well. This expectation has been confirmed in the context of GR by \citet{Sugiyama:2020uil}. \citet{Hirano:2020dom} has shown that when the shift term deviates from $1$, ultra-violet divergence appears in the nonlinear correction term in the power spectrum, i.e. referred to as the $1$-loop term, leading to unattainable converged values. Therefore, we expect similar behavior in the bispectrum and focus only on scales larger than $80\hMpc$, where the loop correction term will not make a significant contribution.

        \vspace{2mm}

    \item  The bin widths are $5\hMpc$ for the 2PCF and $10\hMpc$ for the 3PCF; the 3PCF has a wider bin width than the 2PCF to reduce the number of data bins. These bin widths are the same as those used in \citet{Sugiyama:2020uil} for the anisotropic BAO analysis using the 2PCF and 3PCF.

        \vspace{2mm}

    \item The multipole components of the 2PCF and 3PCF used in the analysis are $\xi_0$, $\xi_2$, $\zeta_{000}$, $\zeta_{110}$, $\zeta_{202}$, and $\zeta_{112}$. In particular, $\zeta_{000}$, $\zeta_{110}$, and $\zeta_{112}$ are only considered for $r_1 \geq r_2$ since $\zeta_{\ell_1\ell_2\ell}(r_1,r_2)=\zeta_{\ell_2\ell_1\ell}(r_2,r_1)$. In this case, the total number of data bins is $202$.

        \vspace{2mm}

    \item The multipole components of the 2PCF and 3PCF are measured using an FFT\footnote{Fast Fourier Transform}-based estimator~\citep{Sugiyama:2018yzo}. The theoretical models for the 2PCF and 3PCF are then computed according to Section $4$ in S23, taking into account the window function effect.

        \vspace{2mm}

    \item The eight parameters constrained in this analysis are $(b_1\sigma_8)$, $(f\sigma_8)$, $(F_{\rm g}\sigma_8)$, $(F_{\rm s}\sigma_8)$, $(F_{\rm t}\sigma_8)$, $(G_{\rm g}\sigma_8)$, $E_{\rm s}$, and $(G_{\rm t}\sigma_8)$; the constraint on the $E_{\rm s}$ parameter is the main result in this paper. 

        \vspace{2mm}

    \item The AP effect~\citep{Alcock:1979mp} is ignored in our analysis. However, the AP effect can be determined by the 2PCF at a few percent and is not expected to significantly affect the constraint results for the parameters that characterize the nonlinear fluctuations of interest in this paper, such as $E_{\rm s}$.

        \vspace{2mm}

    \item The galaxy data used in the analysis is the final galaxy clustering dataset, Data Release 12~\citep[DR12;][]{Alam:2015mbd} from the Baryon Oscillation Spectroscopic Survey~\citep[BOSS;][]{Eisenstein:2011sa,Bolton:2012hz,Dawson:2012va}. The BOSS survey includes four galaxy samples, CMASS, LOWZ, LOWZ2, and LOWZ3, which are combined into a single sample~\citep{Reid:2015gra}. This combined DR12 sample covers the redshift range $z=0.2-0.75$ and is divided into the two redshift bins, $0.2<z<0.5$ and $0.5<z<0.75$, which have the mean redshifts $z=0.38$ and $z=0.61$, respectively. Furthermore, the DR12 sample is observed across two galactic hemispheres, the Northern and Southern Galactic Caps, called NGC and SGC respectively. Thus, the four galaxy samples considered in our analysis are NGC at $z=0.38$, SGC at $z=0.38$, NGC at $z=0.61$, and SGC at $z=0.61$.

        \vspace{2mm}

    \item The 2PCF and 3PCF covariance matrices are computed by measuring the 2PCF and 3PCF from the publicly available $2048$ MultiDark-Patchy mock catalogues~\citep[Patchy mocks;][]{Klypin:2014kpa,Kitaura:2015uqa}.

        \vspace{2mm}

    \item For the NGC and SGC galaxy samples at $z=0.38$, the $p$-values calculated from the best parameter values obtained by our analysis are less than $0.05$, indicating that the theoretical 3PCF model does not fit the measurements well. The fact that such discrepancies between the data and the model occur even in a general parameter space suggests that this is likely to be an indication of systematics. Unfortunately, the reason for this cannot be identified in this paper. Therefore, following Section $8$ in S23, we multiplied the 3PCF covariance matrices measured from the NGC and SGC at $z = 0.38$ by a phenomenological pre-factor of $1.15$ and $1.25$, respectively, to increase the final $p$-value obtained. However, we found that this manipulation had little effect on the final $E_{\rm s}$-constraint. This suggests that the degeneracy between the parameters is the main limitation of our analysis, rather than the $15 - 25\%$ changes in the 3PCF covariance matrices.

        \vspace{2mm}

    \item The Hartlap~\citep[Eq.~(17) in][]{Hartlap:2006kj} and $M_1$~\citep[Square root of Eq.~(18) in][]{Percival:2013sga} factors are used to correct for the effect of errors in the covariance matrix, computed from a finite number of mock catalogues, on the final parameter errors. The $M_2$ factor~\citep[square root of Eq.~(22) in][]{Percival:2013sga}, obtained by combining the Hartlap and $M_1$ factors, is $M_2 = 1.105$ in our analysis, close enough to $1$ for conservative data analysis.

        \vspace{2mm}

    \item The flat prior distribution of the parameter of interest is determined based on the error from a Fisher analysis, performed in the same setting as the main analysis. The fiducial parameter value $\theta_{\rm fid}$, assumed in performing the Fisher analysis, is calculated from the cosmological parameters introduced in Section~\ref{Sec:Introduction} and the linear bias parameter $b_1=2$. With the standard deviation of the parameters obtained by the Fisher analysis being $\sigma_{\rm fisher}(\theta)$, then $\theta_{\rm fid}\pm 5\,\sigma_{\rm fid}(\theta)$ is used as the flat prior distribution.

        \vspace{2mm}

    \item The likelihood of the parameters is computed using the Markov Chain Monte Carlo (MCMC) algorithm implemented in \textsc{Monte Python}~\citep{Brinckmann:2018cvx}. We ensure the convergence of each MCMC chain by imposing $R-1 \lesssim {\cal O}(10^{-4})$, where $R$ is the standard Gelman-Rubin criteria~\citep{Gelman:1992StaSc...7..457G}. 

        The convergence of the results is also checked through the following method. First, eight independent MCMC chains are generated, and the mean and standard deviation of the $E_{\rm s}$ parameter, $(E_{\rm s})_{\rm mean}$ and $(E_{\rm s})_{\rm std}$, are calculated from each chain. Next, the standard deviation of the mean, $((E_{\rm s})_{\rm mean})_{\rm std}$, and the mean of the standard deviation, $((E_{\rm s})_{\rm std})_{\rm mean}$, are calculated from the eight mean values and standard deviations. Finally, the ratio $((E_{\rm s})_{\rm mean})_{\rm std}/((E_{\rm s})_{\rm std})_{\rm mean}$ is checked to be less than $10\%$. Our final $E_{\rm s}$ constraint is obtained by combining all eight chains into a single chain.

\end{enumerate}

\section{Results}
\label{Sec:Results}

\begin{figure*}
    \centering
    \scalebox{0.95}{\includegraphics[width=\textwidth]{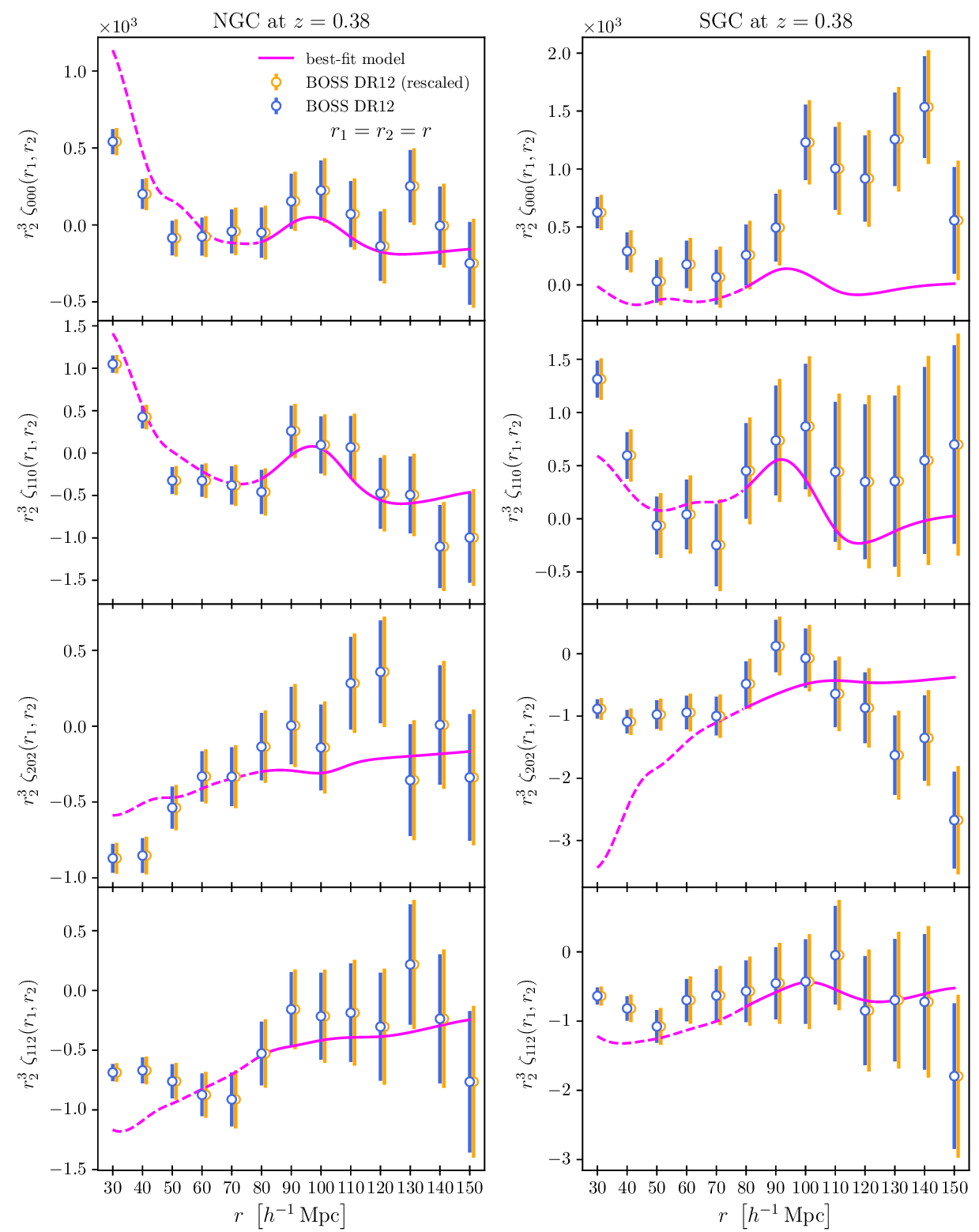}}
    \caption{Multipole components of the 3PCF, i.e. $\zeta_{000}$, $\zeta_{110}$, $\zeta_{202}$ and $\zeta_{112}$, measured from the NGC and SGC samples at $z=0.38$ (blue points). For the sake of simplicity, these plots are shown as a function of $r_1=r_2=r$, even though the actual MCMC analysis also uses the case $r_1\neq r_2$. The error bars are the standard deviation of the 3PCF measurements computed from $2048$ Patchy mocks. The orange error bars are the rescaled ones described in (x) of Section~\ref{Sec:Analysis}, which are used in the MCMC analysis. Also plotted are the theoretical models computed from the best-fit parameter values obtained from the MCMC analysis (magenta lines); they are shown as solid lines at the scales $r\geq 80\hMpc$ used in the analysis, and as dashed lines at smaller scales.}
    \label{fig:3PCF_zbin1}
\end{figure*}

\begin{figure*}
    \centering
    \scalebox{0.95}{\includegraphics[width=\textwidth]{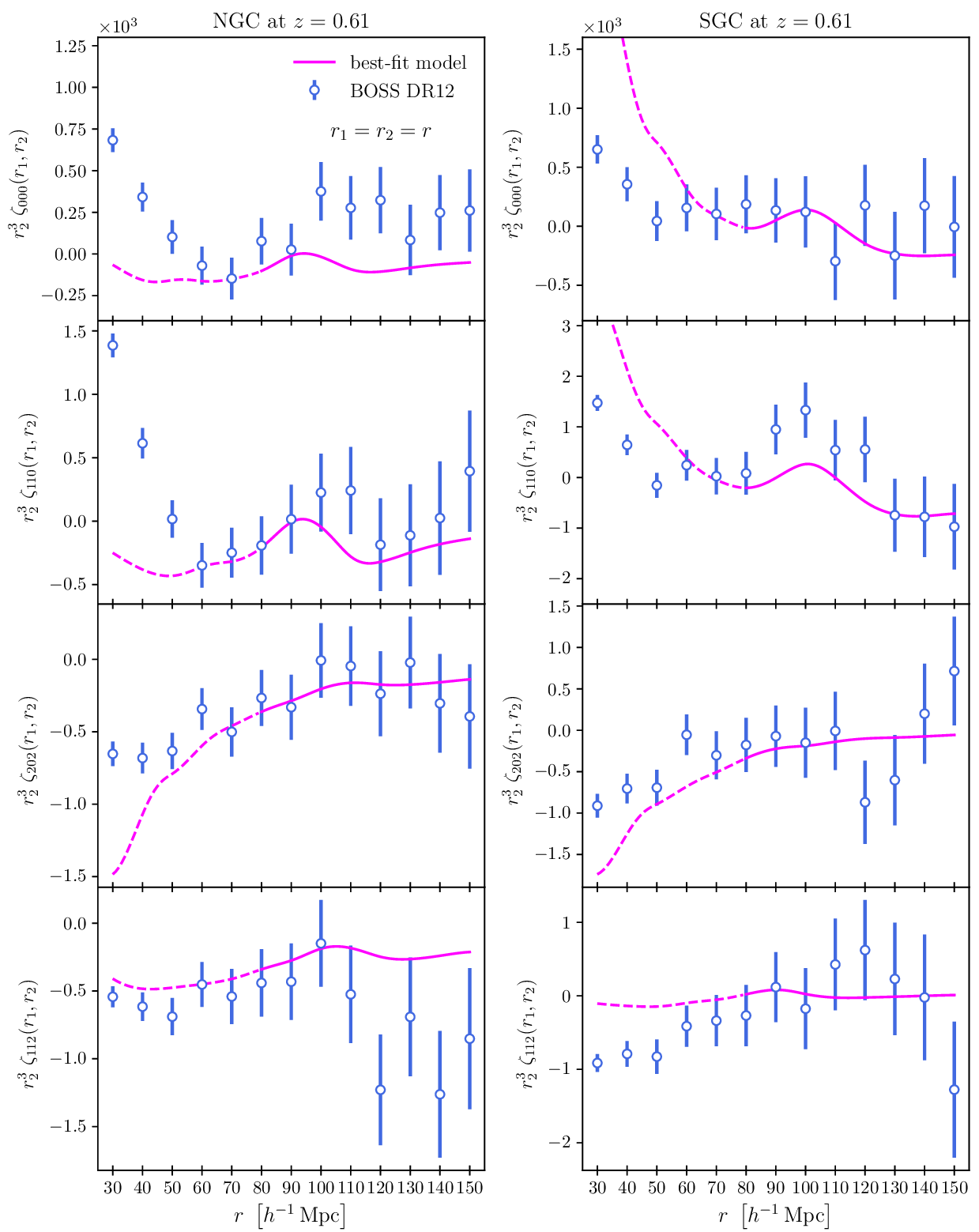}}
    \caption{Same as Figure~\ref{fig:3PCF_zbin1}, except that the results at $z=0.61$ are shown.}
    \label{fig:3PCF_zbin3}
\end{figure*}

\subsection{Measurements}

Figures~\ref{fig:3PCF_zbin1} and \ref{fig:3PCF_zbin3} show the multipole components of the 3PCF measured from the BOSS DR12 galaxies and the corresponding theoretical models calculated with the best-fit parameters in Table~\ref{Table:mean_std}. For the monopole components ($\zeta_{000}$ and $\zeta_{110}$), a BAO peak is expected to appear around $100\hMpc$. For example, $\zeta_{000}$ and $\zeta_{110}$ measured from NGC at $z=0.38$ show a relatively clear BAO signal (see Figure~\ref{fig:3PCF_zbin1}, upper left two panels), but the BAO signal is not seen in some galaxy samples. Also, as noted by S23 and discussed in (x) of Section~\ref{Sec:Analysis}, $\zeta_{000}$ measured from SGC at $z=0.38$ shows statistically significant differences from the theoretical model on large scales (see Figure~\ref{fig:3PCF_zbin1}, upper right panel).

Although the 3PCF multipole, $\zeta_{\ell_1\ell_2\ell}$, is a function of $r_1$ and $r_2$, only the case $r_1=r_2$ is plotted here to simplify the figure; see Figures $12$-$19$ in S23 for the results for $r_1\neq r_2$.

\subsection{Constraints on \texorpdfstring{$E_{\rm s}$}{}}
\label{Sec:Es}

\begin{figure}
    \centering
    \scalebox{1.0}{\includegraphics[width=\columnwidth]{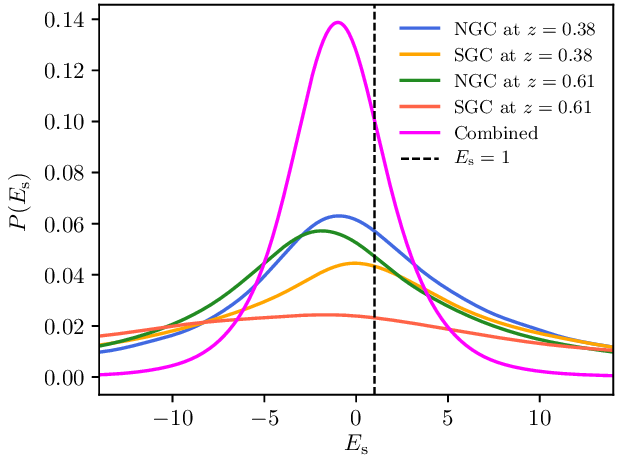}}
    \caption{One-dimensional marginalised posterior probability distributions for $E_{\rm s}$. Results are shown for NGC at $z=0.38$ (blue), SGC at $z=0.38$ (orange), NGC at $z=0.61$ (green), SGC at $z=0.61$ (red), and the four samples combined (magenta). A vertical line with $E_{\rm s}=1$ (black dashed line) is also plotted, indicating that the consistency relation is satisfied.}
    \label{fig:Likelihoods}
\end{figure}

\begin{figure}
    \centering
    \scalebox{1.0}{\includegraphics[width=\columnwidth]{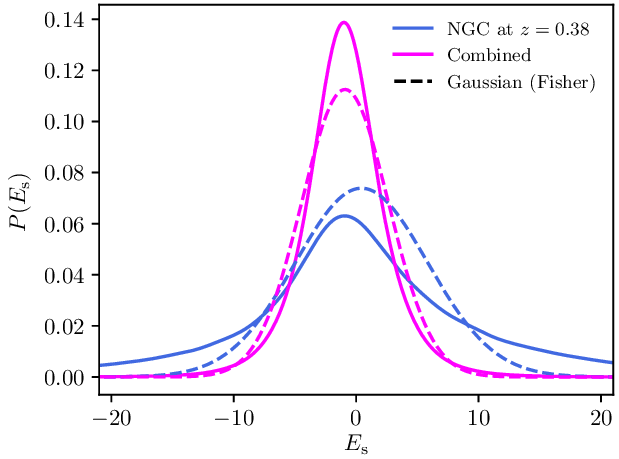}}
    \caption{
        Same as Figure~\ref{fig:Likelihoods}, except that the Gaussian distribution functions assumed in the Fisher analysis are also plotted simultaneously as dashed lines. For clarity of display, only two cases are plotted, NGC at $z=0.38$ and the combined sample. The Gaussian distributions are plotted with the mean value given in the $(E_{\rm s})_{\rm mean}$ column of Table~\ref{Table:results} and the standard deviation predicted by the Fisher analysis input.
    }
    \label{fig:Fisher}
\end{figure}

\begin{table*}
\centering
\begin{tabular}{rrrrrr}
\hline\hline
& $(E_{\rm s})_{\rm mean}$ & $(E_{\rm s})_{\rm std}$ (fisher) & $(E_{\rm s})_{-1\sigma}$ & $(E_{\rm s})_{+1\sigma}$ & $\chi_{\rm min}^2/{\rm DoF}$ ($p$-value)\\
\hline
NGC at $z=0.38$ & $ 0.44$ & $ 9.85\, (5.40)$ & $ -8.93$ & $ 8.32$ & $207.2/194$ $(0.245)$\\
SGC at $z=0.38$ & $-1.69$ & $16.08\, (8.80)$ & $-12.21$ & $15.13$ & $199.3/194$ $(0.382)$\\
NGC at $z=0.61$ & $-1.42$ & $11.89\, (6.95)$ & $ -9.77$ & $ 9.93$ & $216.6/194$ $(0.128)$\\
SGC at $z=0.61$ & $-7.32$ & $23.93\, (11.6)$ & $-19.44$ & $24.78$ & $203.9/194$ $(0.299)$\\
\hline
Combined four samples & $-0.92$ & $3.72\, (3.64)$ & $-3.26$ & $3.13$  & \multicolumn{1}{c}{-} \\
\hline
\end{tabular}
\caption{Means, standard deviations, and $\pm1\sigma$ errors ($68.27\%$ CL) calculated from the $E_{\rm s}$ posteriors shown in Figure~\ref{fig:Likelihoods}. Results are shown for each galaxy sample and for the four samples combined. The standard deviations of the parameters predicted by the Fisher analysis are given in round brackets. The rightmost column shows the reduced $\chi^2$ computed from the best-fit parameter values, and the corresponding $p$-values, where the degrees of freedom (DoF) are $202-8=194$.}
\label{Table:results}
\end{table*}

Figure~\ref{fig:Likelihoods} shows the one-dimensional marginalised posterior probability distributions for $E_{\rm s}$, and Table~\ref{Table:results} summarises the results of constraining $E_{\rm s}$ computed from the posteriors. The results presented in this table show the results of constraining $E_{\rm s}$ separately for the four BOSS samples and a combination of these results.

The $E_{\rm s}$ constraint results from each sample of BOSS galaxies are helpful, for example, in constraining models that vary the coefficient of the shift term from $1$, as presented in Sections~\ref{Sec:DHOST}, \ref{Sec:MassiveNeutrinos}, \ref{Sec:RelativeVelocities}, and \ref{Sec:ExtendedBias}.

On the other hand, from the point of view of examining the violation of the LSS consistency relation, it is also useful to combine all four galaxy samples to see if $E_{\rm s}$ is consistent with $E_{\rm s}=1$. Such an analysis is possible because the predicted $E_{\rm s}$ value for any sample of galaxies and at any redshift is always $E_{\rm s}=1$ if the LSS consistency relation is satisfied. The $E_{\rm s}$ value obtained in this analysis is no longer meaningful as a physical parameter, but is interpreted as a parameter for testing the LSS consistency relation. As a result, we obtain 
\begin{eqnarray}
    E_{\rm s} = -0.92_{-3.26}^{+3.13}
\end{eqnarray}
at the $1\sigma$ level. This result indicates that the present analysis using the BOSS galaxy data does not violate the LSS consistency relation within the statistical error of the data.

\subsection{Comparison with the results of the Fisher analysis}

Table~\ref{Table:results} shows that the errors obtained for each galaxy sample are larger than those predicted by the Fisher analysis, while the combined sample yields a constraint close to the Fisher estimate. This is because the tail of the posterior distribution function obtained for each galaxy sample is more widely spread out than the Gaussian function assumed in the Fisher analysis. On the other hand, when the four galaxy samples are combined, the posterior distribution function approaches the Gaussian function due to the central limit theorem; see Figure~\ref{fig:Fisher} for a comparison of the posterior distribution function of $E_{\rm s}$ and the Gaussian distribution function.

\subsection{Discussions for future research}

In anticipation of future surveys, it is important to note that the magnitude of statistical errors is inversely proportional to the square root of the survey volume. Thus, it naturally follows that the larger the survey volume, the smaller the resulting errors. Currently, the volume of the BOSS data being used in this paper is roughly $4\hV$. By comparison, the survey volume of the Dark Energy Spectroscopic Instrument~\citep[DESI;][]{Aghamousa:2016zmz}\footnote{\url{http://desi.lbl.gov/}} is expected to reach $\sim 40\hV$, which is ten times that of BOSS. Moreover, it is projected that by combining various galaxy surveys, such as Euclid~\citep{Laureijs:2011gra}\footnote{\url{www.euclid-ec.org}} and the Subaru Prime Focus Spectrograph ~\citep[PFS;][]{Takada:2012rn}\footnote{\url{https://pfs.ipmu.jp/index.html}}, we can anticipate an improvement in the current constraint results by a factor of $3\mathchar`-4$.

Furthermore, as demonstrated by \citet{Sugiyama2020:MNRAS.497.1684S}, the shot noise effect determined by the galaxy number density is crucial in assessing statistical errors in the galaxy bispectrum and 3PCF. For example, in the BOSS case, it was shown in Sections~$5$ and $7$ of S23 that a smaller volume but high-density sample at $z=0.38$ can impose stronger constraints on the nonlinear parameters (\ref{Eq:F_2G_2}) using the 3PCF measurement compared to a larger volume but low-density sample at $z=0.61$. Indeed, in this paper, the $E_{\rm s}$ results obtained in Table~\ref{Table:results} show smaller errors for the $z=0.38$ sample than for the $z=0.61$ case. Although the galaxy number density for BOSS is $\sim 3\times 10^{-4}\oV$, for DESI it can reach up to $\sim 7\times 10^{-4}\oV$, depending on the redshift bin, enabling us to anticipate better constraints on $E_{\rm s}$, beyond the actual volume differences.

The prediction of the constraint results for the nonlinear parameters using information on smaller scales than those used in this paper was carried out by the Fisher analysis in Section $7$ of S23 in the context of DHOST theories. In that case, for example, the coefficient $G_{\rm s}\sigma_8$ of the shift term in the nonlinear velocity field is expected to have $\sim 6$ times better error improvement when using up to $30\hMpc$, compared to our current analysis using scales greater than $80\hMpc$. This dramatic improvement in parameter constraints through the use of small scales serves as a strong motivation to further develop theoretical bispectrum models applicable to smaller scales.

The use of more multipole components than the four multipole components ($\zeta_{000}$, $\zeta_{110}$, $\zeta_{202}$ and $\zeta_{112}$) of the 3PCF used in this paper is also expected to improve the constraint results for $E_{\rm s}$ and the other nonlinear parameters.

Finally, note that although all six nonlinear parameters (\ref{Eq:F_2G_2}) are varied in this paper to consider as general a situation as possible, the number of free parameters to be varied is reduced in many cases when actually constraining the specific models presented in Section~\ref{Sec:Examples}. For example, for the constraints on neutrino masses in Section~\ref{Sec:MassiveNeutrinos}, assuming the standard bias effects in Section~\ref{Sec:StandardBias}, all the growth, shift and tidal terms of the nonlinear velocity field can be used to constrain the neutrino mass. The results will therefore be better than the constraint results in this paper, which use only $E_{\rm s}$.

\section{Conclusions}
\label{Sec:Conclusions}

This paper is the first work to test the consistency relation for the LSS from actual galaxy clustering data. We have made this analysis possible through a joint analysis of anisotropic 2PCFs and 3PCFs measured from the BOSS DR12 galaxy data. While the anisotropic component of the 3PCF (or bispectrum) has mainly been used to improve the results of the 2PCF-only analysis~\citep[e.g.,][]{Sugiyama:2020uil,D'Amico:2022arXiv220608327D,Ivanov:2023arXiv230204414I}, the results of this paper open a new observational window for anisotropic 3PCF analysis.

The LSS consistency relation relates the three-point statistics in the squeezed limit to the two-point statistics. The squeezed limit corresponds to extracting only the shift terms that appear in the second-order density and velocity fluctuations, and the LSS consistency relation is satisfied when the coefficients of the shift terms, denoted $F_{\rm s}$ and $G_{\rm s}$ (\ref{Eq:F_2G_2}), are $F_{\rm s}=G_{\rm s}=1$. Conversely, the LSS consistency relation breaks down when $F_{\rm s}$ and $G_{\rm s}$ deviate from $1$, e.g. due to multi-component fluids, modified gravity, and their associated bias effects. However, among the three symmetric wavenumbers, $k_1$, $k_2$ and $k_3$, on which the bispectrum depends, taking the squeezed limit $k_1\to0$, the dependence of the remaining $k_2$ and $k_3$ becomes exchange symmetric, cancelling the coefficient modifications of the shift terms and behaving as if the LSS consistency relation were satisfied~\citep{Crisostomi:2019vhj}. Furthermore, we pointed out in Section~\ref{Sec:Parameterization} that the coefficients of the shift terms are degenerate with the parameter $\sigma_8$ and appear in the form of $(F_{\rm s}\sigma_8)$ and $(G_{\rm s}\sigma_8)$, so we cannot directly constrain $F_{\rm s}$ and $G_{\rm s}$.

Two crucial ideas for solving the problems in the above paragraph are presented in Sections~\ref{Sec:Parameterization} and~\ref{Sec:ConsistencyRelation}. The first idea is to test the LSS consistency relation independently of $\sigma_8$ by defining the $E_{\rm s}$ parameter (\ref{Eq:ES}) as the ratio of $(G_{\rm s}\sigma_8)$ to $(F_{\rm s}\sigma_8)$ and checking whether $E_{\rm s}$ deviates from $1$. Note that $E_{\rm s}\neq1$ is a sufficient but not a necessary condition for showing a violation of the LSS consistency relation, since there may be theories that satisfy $F_{\rm s}=G_{\rm s}\neq1$. The second idea is to ensure that our results hold in as many different situations as possible, we constrain $E_{\rm s}$ in a general parameter space framework with the coefficients of the growth, tidal, and shift terms as free parameters. Section~\ref{Sec:Examples} provides examples of models that are and are not included in our proposed parameterisation.

This analysis requires information about the nonlinearity of the velocity field, which requires dealing with the anisotropic component of the galaxy three-point statistic caused by the RSD effect. In this paper, we adopt a method of decomposing the anisotropic 3PCF using the TripoSH basis function in Section~\ref{Sec:Bispectrum}. This analysis method has been established in a series of papers by~\citet{Sugiyama:2018yzo,Sugiyama:2020uil,Sugiyama:2023arXiv230206808S}. In particular, our analysis method is similar to the one used in \citet{Sugiyama:2023arXiv230206808S} to test DHOST theories from BOSS galaxies, except for the different parameters treated. Therefore, those interested in learning more about the analysis methods discussed in Section~\ref{Sec:Analysis} are referred to that paper.

We have constrained $E_{\rm s}$ from two perspectives using the four galaxy samples from BOSS DR12. The first is a constraint on $E_{\rm s}$ from each galaxy sample that allows a physical interpretation by a specific model, as presented in Section~\ref{Sec:Examples}. The second focuses on the violation of the LSS consistency relation and examines whether $E_{\rm s}$ deviates from $1$ using the combined four samples. In this case, $E_{\rm s}$ is no longer interpreted as a physical parameter but as a parameter for testing the LSS consistency relation. In both cases, the results are consistent with $E_{\rm s}=1$ within the $1\sigma$ error, as shown in Table~\ref{Table:results}. In particular, in the second case we obtained $E_{\rm s} = -0.92_{-3.26}^{+3.13}$. The results of this paper indicate that the LSS consistency relation is not violated within the statistical errors of the data in this analysis using the BOSS galaxy data.

In the future, several extensions can be made. First, it should be possible to include more 3PCF multipole components in the analysis. Second, the AP effect should be included in the analysis, and degeneracy relations between parameters with the AP effect should be considered. Finally, an attempt should be also made to improve the theoretical model of the 3PCF to include information at smaller scales. While attempting these improvements, the present analysis can be directly applied to upcoming spectroscopic galaxy surveys, such as DESI, Euclid and PFS.

\section*{Acknowledgements}

%NSS acknowledges financial support from JSPS KAKENHI Grant Number 19K14703. Numerical computations were carried out on Cray XC50 at Center for Computational Astrophysics, National Astronomical Observatory of Japan. 
%SS acknowledges support for this work from NSF-2219212.
%SS is also supported by World Premier International Research Center Initiative (WPI Initiative), MEXT, Japan.
%This project has received funding from the European Research Council (ERC) under the European Union's Horizon 2020 research and innovation program (grant agreement 853291). FB is a University Research Fellow.

NSS acknowledges financial support from JSPS KAKENHI Grant Number 19K14703. Numerical computations were carried out on Cray XC50 at Center for Computational Astrophysics, National Astronomical Observatory of Japan. 
%The work of SH was supported by JSPS KAKENHI Grants No.~JP21H01080.
The work of TK was supported by
JSPS KAKENHI Grant No.~JP20K03936 and
MEXT-JSPS Grant-in-Aid for Transformative Research Areas (A) ``Extreme Universe'',
No.~JP21H05182 and No.~JP21H05189.
The work of DY was supported in part by JSPS KAKENHI Grants No.~19H01891, No.~22K03627.
SS acknowledges the support for this work from NSF-2219212. 
SS is supported in part by World Premier International Research Center Initiative (WPI Initiative), MEXT, Japan.
H-JS is supported by the U.S. Department of Energy, Office of Science, Office of High Energy Physics under DE-SC0019091 and DE-SC0023241.
This project has received funding from the European Research Council (ERC) under the European Union’s Horizon 2020 research and innovation program (grant agreement 853291). FB is a University Research Fellow.

%%%%%%%%%%%%%%%%%%%%%%%%%%%%%%%%%%%%%%%%%%%%%%%%%%

\section*{Data Availability}

The data underlying this article are available at the SDSS database (\url{https://www.sdss.org/dr12/}).
A complete set of codes for data analysis is packaged under the name \textsc{HITOMI} and is publicly available at \url{https://github.com/naonori/hitomi.git}.



\bibliographystyle{mnras}
\bibliography{ms} % if your bibtex file is called example.bib

%%%%%%%%%%%%%%%%% APPENDICES %%%%%%%%%%%%%%%%%%%%%
\appendix
\section{Other nonlinear parameters}
\label{Sec:OtherParameters}

While the main text focuses only on the results for $E_{\rm s}$, this appendix summarises the results for the other parameters (see (vi) in Section~\ref{Sec:Analysis}) that were varied simultaneously in the MCMC analysis. Results for $G_{\rm s}\sigma_8$ are also reported as $G_{\rm s}\sigma_8 = (F_{\rm s}\sigma_8)\, E_{\rm s}$. Table~\ref{Table:mean_std} shows the best-fit values, means, and standard deviations obtained from the four BOSS samples for the two parameters appearing in linear theory ($b_1\sigma_8$ and $f\sigma_8$) and for the six nonlinear parameters ($F_{\rm g}\sigma_8$, $F_{\rm s}\sigma_8$, $F_{\rm t}\sigma_8$, $G_{\rm g}\sigma_8$, $G_{\rm s}\sigma_8$, $G_{\rm t}\sigma_8$). The covariance matrices for these parameters are shown in Table~\ref{Table:cov}. For illustration, the marginalised one- and two-dimensional posteriors of the parameters are plotted for NGC at $z=0.38$. The results presented in Tables~\ref{Table:mean_std} and \ref{Table:cov} should not only be used to test the LSS consistency relation, which is the subject of this paper but can also be used directly to constrain the various specific models presented in Section~\ref{Sec:Examples}.

In addition to the $E_{\rm s}\neq 1$ condition, it can also be argued that a signal $F_{\rm s}\sigma_8<0$ is a violation of the LSS consistency relation if it is found (see Section~\ref{Sec:ConsistencyRelation}). Therefore, the results on $F_{\rm s}\sigma_8$ from Table~\ref{Table:mean_std} are summarised as follows:
\begin{eqnarray}
    F_{\rm s}\sigma_8 =  \begin{cases} 
            0.715\pm0.685 & (\mbox{NGC at $z=0.38$}) \\
            0.656\pm 1.62 & (\mbox{SGC at $z=0.38$}) \\
            0.883\pm0.845 & (\mbox{NGC at $z=0.61$}) \\
            0.612\pm1.18  & (\mbox{SGC at $z=0.61$}) 
        \end{cases} .
\end{eqnarray}
As shown above, since $F_{\rm s}\sigma_8<0$ cannot be statistically significant, it can be concluded that no violation of the LSS consistency relation was found in the current analysis.

\begin{table*}
\centering
\begin{tabular}{crrrrrrrr}
\hline
\hline
\multicolumn{9}{c}{NGC at $z=0.38$} \\
\hline
& $b_1\sigma_8$ & $f \sigma_8$ & $F_{\rm g}\sigma_8$ & $F_{\rm s}\sigma_8$ & $F_{\rm t}\sigma_8$ & $G_{\rm g}\sigma_8$ & $G_{\rm s}\sigma_8$ & $G_{\rm t}\sigma_8$\\
\hline
best-fit & $1.35$ & $0.469$ & $1.20$ & $0.753$ & $-0.109$ & $-1.80$ & $-0.0987$ & $2.00$\\
mean & $1.23$ & $0.433$ & $1.32$ & $0.715$ & $-0.0191$ & $-2.47$ & $0.53$ & $2.75$ \\
std. & $0.18$ & $0.107$ & $0.715$ & $0.685$ & $0.452$ & $2.54$ & $4.43$ & $2.08$ \\
\hline                
\vspace{0.2cm}\\
\hline                
\hline
\multicolumn{9}{c}{SGC at $z=0.38$} \\
\hline
& $b_1\sigma_8$ & $f \sigma_8$ & $F_{\rm g}\sigma_8$ & $F_{\rm s}\sigma_8$ & $F_{\rm t}\sigma_8$ & $G_{\rm g}\sigma_8$ & $G_{\rm s}\sigma_8$ & $G_{\rm t}\sigma_8$\\
\hline
best-fit & $1.19$ & $0.569$ & $0.142$ & $1.65$ & $-1.25$ & $5.58$ & $2.12$ & $-0.613$ \\
mean & $0.627$ & $0.681$ & $0.907$ & $0.656$ & $-0.47$ & $4.35$ & $4.93$ & $-1.36$ \\
std. & $0.316$ & $0.263$ & $2.38$ & $1.62$ & $1.37$ & $4.71$ & $12.7$ & $5.34$ \\
\hline                
\vspace{0.2cm}\\
\hline
\hline
\multicolumn{9}{c}{NGC at $z=0.61$} \\
\hline
& $b_1\sigma_8$ & $f \sigma_8$ & $F_{\rm g}\sigma_8$ & $F_{\rm s}\sigma_8$ & $F_{\rm t}\sigma_8$ & $G_{\rm g}\sigma_8$ & $G_{\rm s}\sigma_8$ & $G_{\rm t}\sigma_8$\\
\hline
best-fit & $1.27$ & $0.366$ & $0.107$ & $0.911$ & $-0.191$ & $3.24$ & $-0.882$ & $0.00682$ \\
mean & $1.08$ & $0.361$ & $-0.0782$ & $0.883$ & $0.0875$ & $2.91$ & $-0.633$ & $0.393$ \\
std. & $0.158$ & $0.109$ & $0.868$ & $0.845$ & $0.574$ & $3.05$ & $7.85$ & $2.7$ \\
\hline                
\vspace{0.2cm}\\
\hline
\hline
\multicolumn{9}{c}{SGC at $z=0.61$} \\
\hline
& $b_1\sigma_8$ & $f \sigma_8$ & $F_{\rm g}\sigma_8$ & $F_{\rm s}\sigma_8$ & $F_{\rm t}\sigma_8$ & $G_{\rm g}\sigma_8$ & $G_{\rm s}\sigma_8$ & $G_{\rm t}\sigma_8$\\
\hline
best-fit & $1.27$ & $0.266$ & $1.45$ & $-0.294$ & $1.37$ & $5.34$ & $5.96$ & $-9.76$ \\
mean & $0.943$ & $0.312$ & $1.41$ & $0.612$ & $1.73$ & $3.8$ & $0.0212$ & $-6.57$ \\
std. & $0.235$ & $0.168$ & $1.93$ & $1.18$ & $1.12$ & $6.81$ & $25.3$ & $4.24$ \\
\hline
\hline\end{tabular}         
\caption{Best-fit values, means, and standard deviations for $(b_1\sigma_8)$, $(f\sigma_8)$, $(F_{\rm g}\sigma_8)$, $(F_{\rm s}\sigma_8)$, $(F_{\rm t}\sigma_8)$, $(G_{\rm g}\sigma_8)$, $(G_{\rm s}\sigma_8)$ and $(G_{\rm t}\sigma_8)$ obtained in the joint analysis of the 2PCF and the 3PCF using the four BOSS samples. The results for $(G_{\rm s}\sigma_8)= (F_{\rm s}\sigma_8)\, E_{\rm s}$ have been obtained from the MCMC chain of $E_{\rm s}$ and $(F_{\rm s}\sigma_8)$.
}
\label{Table:mean_std}
\end{table*}

\begin{table*}
\centering
\begin{tabular}{crrrrrrrr}
\hline
\hline
\multicolumn{9}{c}{NGC at $z=0.38$} \\
\hline
& $b_1\sigma_8$ & $f \sigma_8$ & $F_{\rm g}\sigma_8$ & $F_{\rm s}\sigma_8$ & $F_{\rm t}\sigma_8$ & $G_{\rm g}\sigma_8$ & $G_{\rm s}\sigma_8$ & $G_{\rm t}\sigma_8$\\
\hline
$b_1\sigma_8$       & $0.0318$ & $-0.0109$ & $-0.0192$ & $-0.028$ & $0.00321$ & $-0.0864$ & $-0.161$ & $-0.0477$     \\
$f \sigma_8$        & $-0.0109$ & $0.0112$ & $0.00192$ & $0.00325$ & $-0.000899$ & $0.0823$ & $-0.0241$ & $-0.0288$  \\
$F_{\rm g}\sigma_8$ & $-0.0192$ & $0.00192$ & $0.521$ & $0.0496$ & $0.151$ & $-0.968$ & $-0.425$ & $-0.103$          \\ 
$F_{\rm s}\sigma_8$ & $-0.028$ & $0.00325$ & $0.0496$ & $0.593$ & $-0.236$ & $0.155$ & $-1.02$ & $0.404$             \\ 
$F_{\rm t}\sigma_8$ & $0.00321$ & $-0.000899$ & $0.151$ & $-0.236$ & $0.22$ & $-0.313$ & $0.0416$ & $-0.327$         \\ 
$G_{\rm g}\sigma_8$ & $-0.0864$ & $0.0823$ & $-0.968$ & $0.155$ & $-0.313$ & $6.26$ & $3.64$ & $-1.58$               \\ 
$G_{\rm s}\sigma_8$ & $-0.161$ & $-0.0241$ & $-0.425$ & $-1.02$ & $0.0416$ & $3.64$ & $20.4$ & $-4.52$               \\ 
$G_{\rm t}\sigma_8$ & $-0.0477$ & $-0.0288$ & $-0.103$ & $0.404$ & $-0.327$ & $-1.58$ & $-4.52$ & $4.09$             \\ 
\hline
\vspace{0.3cm}\\
\hline
\hline
\multicolumn{9}{c}{SGC at $z=0.38$} \\
\hline
& $b_1\sigma_8$ & $f \sigma_8$ & $F_{\rm g}\sigma_8$ & $F_{\rm s}\sigma_8$ & $F_{\rm t}\sigma_8$ & $G_{\rm g}\sigma_8$ & $G_{\rm s}\sigma_8$ & $G_{\rm t}\sigma_8$\\
\hline
$b_1\sigma_8$       & $0.0865$ & $-0.0422$ & $0.0464$ & $0.0969$ & $-0.0934$ & $-0.0671$ & $-0.279$ & $0.0788$   \\
$f \sigma_8$        & $-0.0422$ & $0.0631$ & $-0.0558$ & $-0.0621$ & $0.0296$ & $0.00688$ & $-0.404$ & $-0.202$  \\
$F_{\rm g}\sigma_8$ & $0.0464$ & $-0.0558$ & $4.51$ & $0.296$ & $0.435$ & $-3.5$ & $-0.245$ & $0.0227$           \\ 
$F_{\rm s}\sigma_8$ & $0.0969$ & $-0.0621$ & $0.296$ & $3.31$ & $-0.516$ & $0.433$ & $-3.37$ & $-0.0595$         \\ 
$F_{\rm t}\sigma_8$ & $-0.0934$ & $0.0296$ & $0.435$ & $-0.516$ & $1.99$ & $-0.549$ & $-0.63$ & $-1.27$          \\ 
$G_{\rm g}\sigma_8$ & $-0.0671$ & $0.00688$ & $-3.5$ & $0.433$ & $-0.549$ & $22.5$ & $12.6$ & $-11.5$            \\ 
$G_{\rm s}\sigma_8$ & $-0.279$ & $-0.404$ & $-0.245$ & $-3.37$ & $-0.63$ & $12.6$ & $171$ & $-17.9$              \\ 
$G_{\rm t}\sigma_8$ & $0.0788$ & $-0.202$ & $0.0227$ & $-0.0595$ & $-1.27$ & $-11.5$ & $-17.9$ & $27.6$          \\ 
\hline
\vspace{0.3cm}\\
\hline
\hline
\multicolumn{9}{c}{NGC at $z=0.61$} \\
\hline
& $b_1\sigma_8$ & $f \sigma_8$ & $F_{\rm g}\sigma_8$ & $F_{\rm s}\sigma_8$ & $F_{\rm t}\sigma_8$ & $G_{\rm g}\sigma_8$ & $G_{\rm s}\sigma_8$ & $G_{\rm t}\sigma_8$\\
\hline
$b_1\sigma_8$       & $0.024$ & $-0.00748$ & $0.0309$ & $-0.0267$ & $-0.018$ & $-0.0598$ & $-0.0707$ & $-0.0294$   \\
$f \sigma_8$        & $-0.00748$ & $0.0117$ & $-0.0141$ & $-0.00346$ & $0.00278$ & $0.0298$ & $0.0602$ & $0.00315$ \\
$F_{\rm g}\sigma_8$ & $0.0309$ & $-0.0141$ & $0.806$ & $0.0312$ & $0.204$ & $-1.49$ & $-0.814$ & $-0.332$          \\ 
$F_{\rm s}\sigma_8$ & $-0.0267$ & $-0.00346$ & $0.0312$ & $0.894$ & $-0.311$ & $0.00472$ & $-2.62$ & $0.776$       \\ 
$F_{\rm t}\sigma_8$ & $-0.018$ & $0.00278$ & $0.204$ & $-0.311$ & $0.37$ & $-0.317$ & $0.347$ & $-0.705$           \\ 
$G_{\rm g}\sigma_8$ & $-0.0598$ & $0.0298$ & $-1.49$ & $0.00472$ & $-0.317$ & $9.46$ & $6.37$ & $-2.09$            \\ 
$G_{\rm s}\sigma_8$ & $-0.0707$ & $0.0602$ & $-0.814$ & $-2.62$ & $0.347$ & $6.37$ & $43.3$ & $-10.1$              \\ 
$G_{\rm t}\sigma_8$ & $-0.0294$ & $0.00315$ & $-0.332$ & $0.776$ & $-0.705$ & $-2.09$ & $-10.1$ & $7.75$           \\ 
\hline
\vspace{0.3cm}\\
\hline
\hline
\multicolumn{9}{c}{SGC at $z=0.61$} \\
\hline
& $b_1\sigma_8$ & $f \sigma_8$ & $F_{\rm g}\sigma_8$ & $F_{\rm s}\sigma_8$ & $F_{\rm t}\sigma_8$ & $G_{\rm g}\sigma_8$ & $G_{\rm s}\sigma_8$ & $G_{\rm t}\sigma_8$\\
\hline
$b_1\sigma_8$       & $0.0576$ & $-0.0196$ & $0.0903$ & $-0.0844$ & $-0.0668$ & $-0.116$ & $-0.625$ & $-0.171$      \\
$f \sigma_8$        & $-0.0196$ & $0.0298$ & $-0.0581$ & $-0.0324$ & $0.0214$ & $0.135$ & $1.19$ & $-0.0697$        \\
$F_{\rm g}\sigma_8$ & $0.0903$ & $-0.0581$ & $3.99$ & $0.155$ & $0.508$ & $-5.36$ & $-4.29$ & $-0.585$              \\ 
$F_{\rm s}\sigma_8$ & $-0.0844$ & $-0.0324$ & $0.155$ & $2.6$ & $-0.453$ & $-0.126$ & $-13.3$ & $1.39$             \\ 
$F_{\rm t}\sigma_8$ & $-0.0668$ & $0.0214$ & $0.508$ & $-0.453$ & $1.14$ & $-0.607$ & $-0.467$ & $-1.1$             \\ 
$G_{\rm g}\sigma_8$ & $-0.116$ & $0.135$ & $-5.36$ & $-0.126$ & $-0.607$ & $44.9$ & $24.5$ & $-6.81$              \\ 
$G_{\rm s}\sigma_8$ & $-0.625$ & $1.19$ & $-4.29$ & $-13.3$ & $-0.467$ & $24.5$ & $651$ & $-33.1$              \\ 
$G_{\rm t}\sigma_8$ & $-0.171$ & $-0.0697$ & $-0.585$ & $1.39$ & $-1.1$ & $-6.81$ & $-33.1$ & $26.6$              \\ 
\hline                
\end{tabular}         
\caption{Covariance matrices for $(b_1\sigma_8)$, $(f\sigma_8)$, $(F_{\rm g}\sigma_8)$, $(F_{\rm s}\sigma_8)$, $(F_{\rm t}\sigma_8)$, $(G_{\rm g}\sigma_8)$, $(G_{\rm s}\sigma_8)$ and $(G_{\rm t}\sigma_8)$ obtained in the joint analysis of the 2PCF and the 3PCF using the four BOSS samples.}            
\label{Table:cov}
\end{table*}

\begin{figure*}
    \centering
    \scalebox{1.0}{\includegraphics[width=\textwidth]{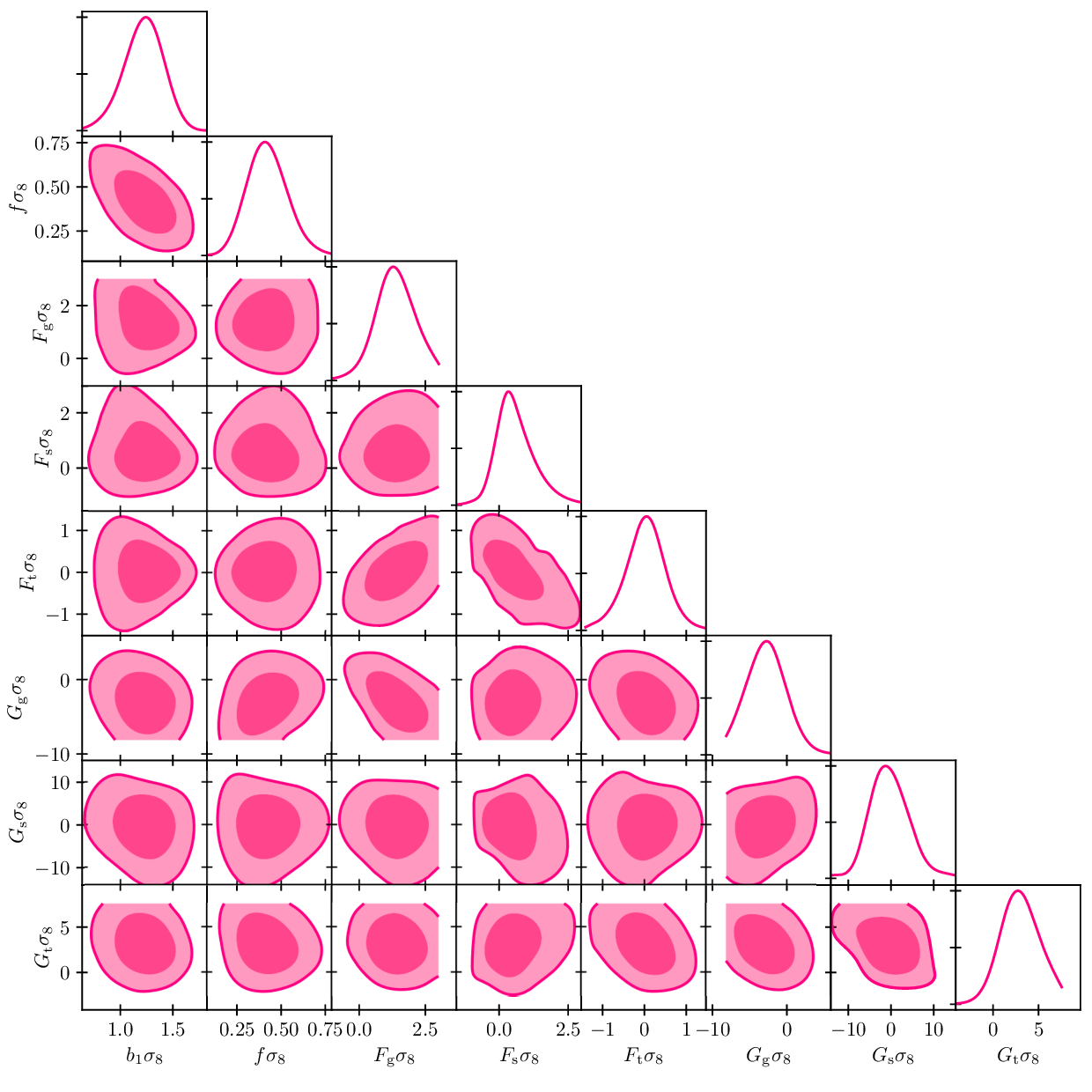}}
    \caption{Marginalized two- and one-dimensional posteriors of the parameters $(b_1\sigma_8)$, $(f\sigma_8)$, $(F_{\rm g}\sigma_8)$, $(F_{\rm s}\sigma_8)$, $(F_{\rm t}\sigma_8)$, $(G_{\rm g}\sigma_8)$, $(G_{\rm s}\sigma_8)$ and $(G_{\rm t}\sigma_8)$. The contours indicate $68.27\%$ and $95.45\%$ confidence levels. The result is for NGC at $z=0.38$.}
    \label{fig:2Dcontour}
\end{figure*}

% Don't change these lines
\bsp	% typesetting comment
\label{lastpage}
\end{document}